\documentclass[noshowpacs,hyper,nofootinbib,amsmath,amssymb,superscriptaddress,11pt]{revtex4}

\pdfoutput=1
\usepackage{subfigure}
\usepackage{graphicx}
\usepackage{latexsym}
\usepackage{amsfonts}
\usepackage{amssymb}
\usepackage{amsmath}
\usepackage{slashed}
\usepackage{array}
\usepackage{feynmp}
\usepackage{url}
\usepackage{multirow}

\makeatletter
\def\l@subsection#1#2{}
\def\l@subsubsection#1#2{}
\makeatother

\allowdisplaybreaks

\newcommand{\be}{\begin{eqnarray}}
\newcommand{\ee}{\end{eqnarray}}
\newcommand{\bea}{\begin{eqnarray}}
\newcommand{\eea}{\end{eqnarray}}

\newcommand{\GeV}{{~\rm GeV}}

\newcommand{\Eslash}{\mbox{${\rm E \kern-0.6em\slash}$}}

\newcommand{\no}\nonumber
\renewcommand{\d}{\rm d}\allowdisplaybreaks

\newcommand{\ZZ}{{\rm Z}^0}

\newcommand{\WW}{{\rm W}^\pm}

\newcommand{\ifb}{{\rm fb}^{-1}}

\newcommand{\gSM}{g_{_{\rm SM}}}
\newcommand{\gDM}{g_{_{\rm DM}}}
\newcommand{\Mmed}{M_{\rm med}}

\begin{document}

\preprint{}

\title{Simplified Models for Dark Matter and Missing Energy Searches at the LHC}


\author{Jalal Abdallah}
\affiliation{Academia Sinica, Institute of Physics, Taiwan}
\author{Adi Ashkenazi}
\affiliation{Department of Physics, Tel Aviv University,Tel Aviv 6139001, Israel}
\author{Antonio Boveia}
\affiliation{Enrico Fermi Institute, University of Chicago, Chicago, IL, USA} 
\author{Giorgio Busoni}
\affiliation{SISSA and INFN, Sezione di Trieste, via Bonomea 265, I-34136 Trieste, Italy}
\author{Andrea De Simone}
\affiliation{SISSA and INFN, Sezione di Trieste, via Bonomea 265, I-34136 Trieste, Italy}
\author{Caterina Doglioni}
\affiliation{Section de Physique, Universit\'e de Gen\`eve,\\
24 quai E. Ansermet, CH-1211 Geneva, Switzerland}
\author{Aielet Efrati}
\affiliation{Department of Particle Physics and Astrophysics, Weizmann Institute of Science, Rehovot 7610001, Israel}
\author{Erez Etzion}
\affiliation{Department of Physics, Tel Aviv University,Tel Aviv 6139001, Israel}
\author{Johanna Gramling}
\affiliation{Section de Physique, Universit\'e de Gen\`eve,\\
24 quai E. Ansermet, CH-1211 Geneva, Switzerland}
\author{Thomas Jacques}
\affiliation{Section de Physique, Universit\'e de Gen\`eve,\\
24 quai E. Ansermet, CH-1211 Geneva, Switzerland}
\author{Tongyan Lin}
\affiliation{Kavli Institute for Cosmological Physics and the Enrico
Fermi Institute, The University of Chicago, 5640 S. Ellis Ave,
Chicago, Il 60637}
\author{Enrico Morgante}
\affiliation{Section de Physique, Universit\'e de Gen\`eve,\\
24 quai E. Ansermet, CH-1211 Geneva, Switzerland}
\author{Michele Papucci}
\affiliation{Berkeley Center for Theoretical Physics, University of California, Berkeley, CA 94720, USA}
\affiliation{Theoretical Physics Group, Lawrence Berkeley National Laboratory, Berkeley, CA 94720, USA}
\author{Bjoern Penning}
\affiliation{Enrico Fermi Institute, University of Chicago, Chicago, IL, USA}
\affiliation{Fermi National Accelerator Laboratory, Batavia, Illinois 60510, USA}
\author{Antonio Walter Riotto}
\affiliation{Section de Physique, Universit\'e de Gen\`eve,\\
24 quai E. Ansermet, CH-1211 Geneva, Switzerland}
\author{Thomas Rizzo}
\affiliation{SLAC National Accelerator Laboratory, Menlo Park, CA, 94025, USA}
\author{David Salek}
\affiliation{Nikhef and GRAPPA, Amsterdam, Netherlands}
\author{Steven Schramm}
\affiliation{Department of Physics, University of Toronto, Toronto, Ontario, Canada M5S1A7}
\author{Oren Slone}
\affiliation{Department of Physics, Tel Aviv University,Tel Aviv 6139001, Israel}
\author{Yotam Soreq}
\affiliation{Department of Particle Physics and Astrophysics, Weizmann Institute of Science, Rehovot 7610001, Israel}
\author{Alessandro Vichi}
\affiliation{Berkeley Center for Theoretical Physics, University of California, Berkeley, CA 94720, USA}
\affiliation{Theoretical Physics Group, Lawrence Berkeley National Laboratory, Berkeley, CA 94720, USA}
\author{Tomer Volansky}
\affiliation{Department of Physics, Tel Aviv University,Tel Aviv 6139001, Israel}
\author{Itay Yavin}
\affiliation{Perimeter Institute for Theoretical Physics, Waterloo, Ontario, Canada}
\affiliation{Department of Physics, McMaster University,  
Hamilton, ON, Canada}
\email[Corresponding author: ]{iyavin@perimeterinstitute.ca}
\author{Ning Zhou}
\affiliation{Department of Physics and Astronomy, University of California, Irvine, California, USA}
\author{Kathryn Zurek}
\affiliation{Berkeley Center for Theoretical Physics, University of California, Berkeley, CA 94720, USA}
\affiliation{Theoretical Physics Group, Lawrence Berkeley National Laboratory, Berkeley, CA 94720, USA}

\newpage
\clearpage

\begin{abstract}
The study of collision events with missing energy as searches for the dark matter (DM) component of the Universe are an essential part of the extensive program looking for new physics at the LHC. Given the unknown nature of DM, the interpretation of such searches should be made broad and inclusive. This report reviews the usage of simplified models in the interpretation of missing energy searches. We begin with a brief discussion of the utility and limitation of the effective field theory approach to this problem. The bulk of the report is then devoted to several different simplified models and their signatures, including $s$-channel and $t$-channel processes. A common feature of simplified models for DM is the presence of additional particles that mediate the interactions between the Standard Model and the particle that makes up DM. We consider these in detail and emphasize the importance of their inclusion as final states in any coherent interpretation. We also review some of the experimental progress in the field, new signatures, and other aspects of the searches themselves. We conclude with comments and recommendations regarding the use of simplified models in Run-II of the LHC. 
\end{abstract}

\maketitle

\newpage
\clearpage

\setcounter{tocdepth}{0}

\tableofcontents

\newpage
\clearpage
\section{Introduction}

Many well-motivated extensions of the Standard Model (SM), such as the Minimal Supersymmetric SM (MSSM)~\cite{Chung:2003fi}, Large extra dimensions (LED)~\cite{ArkaniHamed:1998rs}, little Higgs models with T-parity~\cite{Schmaltz:2005ky}, etc. predict large missing energy signals at high-energy colliders. Often, the production of new particles associated with these extensions of the SM results in more than just missing energy signature, as is the case in the MSSM with the production of squarks and sleptons which cascade decay to the lightest supersymmetric partner (LSP). Such unusual events, with energetic leptons, jets, and large amounts of missing energy contain several discriminating features as compared with the SM and form the basis for powerful searches for new physics (see~\cite{Aad:2014pda} for a recent example). At the same time, there are good reasons to develop searches that do not rely on extra discriminating features aside from large missing energy. Such searches, where large missing energy is the dominant signature of new physics, are the principal subject of this report.

The first strong motivation for missing energy searches is that the models mentioned above allow for the possibility of producing missing energy without it being accompanied by other unusual objects. For example in LED scenarios where energy escapes to the extra dimensions, or in the MSSM where direct pair production of the LSP results only in missing energy. The second reason is the overwhelming evidence for Dark Matter (DM) in the universe. If DM is a new fundamental particle, and if it interacts weakly but not too weakly with the SM, then the annihilation of SM particles into DM constitutes a new source of missing energy in colliders. This picture becomes particularly compelling in light of the WIMP(less) miracle, which connects its mass, coupling, and relic abundance (see e.g.~\cite{Feng:2008ya}). The third reason is that such searches at colliders have a much more broad interpretation, and are sensitive to much more than just stable new particles. Such searches are sensitive also to any new, weakly interacting particle with a lifetime that exceeds about a microsecond since these would leave the detector before depositing their energy. It is therefore well worth the effort to develop a comprehensive search strategy to look for events with large missing energy as their dominant  discriminating signature. 

In hadron colliders the observable quantity associated with undetected particles is of course only the momentum imbalance in the direction transverse to the beam, or the missing transverse momentum\footnote{Here and in what follows we will often abuse the terminology slightly and refer to searches utilizing this imbalance more generally as missing energy searches.}. The magnitude of the missing transverse momentum is known as the missing transverse energy (MET). The simplest and best-known example of a search for large MET is the monojet search looking for a single QCD jet recoiling against nothing. In the underlying particle model the jet is typically assumed to be radiated from the initial state partons in the event before the collision produced the invisible components. It is now common to also include (or at least, not to exclude) multijet events recoiling against MET in the search for missing energy~\cite{Aad:2011xw,Chatrchyan:2012me,ATLAS:2012ky, Abdallah:2013jma,CMS-PAS-EXO-12-048}. If nothing else this is useful because the probability of radiating a second jet from the initial state partons is large at LHC energies (discussed in e.g.~\cite{Haisch:2013ata}). In addition, as we discuss below, in some regions of the parameter space the underlying theoretical models often predict comparable signal in multijet events with missing energy as in the monojet signal. Finally, there are good reasons to consider MET recoil against objects other than QCD jets, such as mono-photon, mono-W, mono-Z, top- and bottom-tagged jets, and mono-lepton searches. 

 Thus, missing energy signatures form a very wide net with which weakly interacting particles, not necessarily forming the dominant component of DM, can be efficiently searched for.  The inclusive nature of these searches calls for the construction of equally broad theoretical models that can be used to interpret the experimental results in a comprehensive fashion. Over the past several years the Effective Field Theory (EFT) approach has gained in popularity since it allows one to focus on a minimal number of degrees-of-freedom, for example the initial partons involved in the reaction (quarks and gluons) and the DM candidate~\cite{Cao:2009uw, Goodman:2010yf,Goodman:2010ku,Bai:2010hh, Cheung:2010ua, Zheng:2010js, Cheung:2011nt, Yu:2011by, Fan:2010gt, Fitzpatrick:2012ix}. It remains agnostic about heavier particles that may be present in a fully renormalizable model and thus allows for a fairly model-independent interpretation. However, as was recognized early on~\cite{Bai:2010hh, Fox:2011fx, Fox:2011pm, Shoemaker:2011vi, Fox:2012ee, Aaltonen:2012jb, Weiner:2012cb}, and more recently in a quantitative way~\cite{Busoni:2013lha, Buchmueller:2013dya, Busoni:2014sya}, the validity of this approach is often questionable at LHC energies where the momentum transfer involved in the reactions is comparable to the scale of non-renormalizable operator being constrained. In other words, the degrees-of-freedom that were assumed to generate these operators are important (in the parlance of EFTs, they should be "integrated-in"). The question then arises: \textit{how do we amend the EFT approach and incorporate the effects of these other particles in the modeling of missing energy searches while continuing to work in a broad and inclusive theoretical framework?} 

Simplified models~\cite{ArkaniHamed:2007fw,Alwall:2008ag,Alves:2011wf} offer a powerful approach to address this issue by including in a minimal model the extra particles and interactions needed to reproduce the non-renormalizable operators. This should not be viewed as a step backwards. On the contrary: as is well-known from other studies, simplified models allow us to focus on the salient kinematical features of a process while ignoring differences among models (such as helicity structure) that LHC measurements are anyways only weakly sensitive to. The DM-EFT operators are a case in point as many of the operators considered (e.g. ${\bar q}\gamma_\alpha q ~{\bar \chi}\gamma^\alpha \chi$ and ${\bar q}\gamma_5\gamma_\alpha q ~{\bar \chi}\gamma_5\gamma^\alpha \chi$) yield similar kinematical distributions at the LHC. Of course, these different operators yield very different behavior in direct- and indirect-detection experiments as we discuss below, but that is not pertinent for the purpose of presenting results from searches at the LHC\footnote{However, it is important that the equivalency of these different choices for LHC phenomenology is clearly communicated so as to avoid misunderstandings with regard to the relevancy of the LHC results to other model choices and other experiments.}. Simplified models also bring to a sharp focus the importance of other searches at the LHC such as multi-jet+MET searches, which can provide complementary bounds on the underlying model. This is so because the new degrees-of-freedom included in the simplified model (which we henceforth refer to as \textit{mediators}) can be produced on-shell and contribute significantly to processes other than the original ones considered within the EFT context. In sections~\ref{sec:qqchichi_s}, \ref{sec:qqchichi_t}, and \ref{sec:ggchichi} we introduce and discuss the different simplified models. 

Alongside missing energy searches at colliders, efforts for direct and indirect detection of DM offer complementary fronts where DM can be searched for (see for example ref.~\cite{Salati:2007zz}). One of the advantages of the EFT approach is that it allows for a straightforward comparison of constraints coming from the different fronts. Simplified models maintain this advantage and allow for an equally straightforward comparison with direct detection experiments as was demonstrated for example in refs.~\cite{An:2012va,An:2013xka,Papucci:2014iwa,Buchmueller:2014yoa}. At the same time, simplified models avoid the pitfalls of the EFT approach by correctly modeling the weaker constraints on models with light mediators. We discuss these points further in the different sections where the simplified models are introduced.

It is the purpose of this report to carefully examine the case for using simplified-models in DM searches at the LHC, and to make recommendations for their adoption in future analyses. It is organized as follows: we begin in section~\ref{sec:problems_with_EFT} with a brief review of the literature and results pertaining to the validity of the EFT approach - much has already been done and so we concentrate on a small set of illustrative examples including both a qualitative as well as a quantitative discussion of the problems that arise. We follow with sections~\ref{sec:qqchichi_s}, \ref{sec:qqchichi_t}, \ref{sec:ggchichi}, and~\ref{sec:third_generation} presenting and discussing a set of simplified-models to be used in LHC DM searches - models with $s$-channel and $t$-channel contributions with different spin assignments for the mediators. In section~\ref{sec:mediators} we also discuss searches for the mediators themselves as well as some of the experimental aspects of current and planned searches in section~\ref{sec:exp_aspects}. We devote section~\ref{sec:recommendations} to general comments and recommendations for the use and presentation of simplified models in Run-II of the LHC and we conclude in section~\ref{sec:conclusions}.


\section{The Problems with the EFT approach}
\label{sec:problems_with_EFT}

The basic idea behind the EFT approach for searches of weakly interacting massive particles (WIMPs) is to consider a set of non-renormalizable operators that couple the partons (quarks and gluons) to a field that represents the WIMP, such as
\be
\mathcal{L}_{\rm EFT} = \frac{1}{\Lambda^2}\left(\bar{q} q\right)\left(\bar{\chi}  \chi\right) \, .
\ee
Here we model the WIMP as a fermion $\chi$ and the quarks are labeled by $q$. Since this is a dimension six operator, we introduced the mass scale $\Lambda$ as some high-energy scale associated with this interaction. Table 1 of ref.~\cite{Goodman:2010ku}, for example, provides a fairly exhaustive list of such operators.  Operators of this kind can simultaneously describe a variety of physics processes: annihilation of a pair of quarks into WIMPs; scattering of WIMPs on quarks; and the annihilation of a WIMP pair into quarks. The calculation of physical processes with such non-renormalizable operators is done in perturbation theory through the energy expansion: processes associated with this operator generally scale as $E^n/\Lambda^n$ where $n$ is some positive power and $E$ is the characteristic energy of the process. This treatment is consistent and the energy expansion is meaningful as long as the energy of the process is small compared with the scale $\Lambda$. 

Therefore the  necessary condition for the EFT approach to provide a valid description is to have a clear separation 
of scales:  the energy scale of the process one wants to describe must be much lower than
 the scale of the underlying microscopic  interactions. 
 In  the context of DM searches, there are several situations where the EFT approach is absolutely
 solid. In indirect searches, for example, 
the energy scale for the non-relativistic annihilation of DM particles 
in the halo is  of the order of the DM mass $m_{\rm DM}$;
for  direct DM searches,   the scattering of DM particles with heavy nuclei occur at energy scales of the order of 10 keV.
Therefore, assuming the mediator is not lighter than $\mathcal{O}(10)$ keV ($\mathcal{O}(m_{\rm DM})$), it is certainly possible to describe processes relevant to (in)direct-detection by means of an EFT
(see e.g. Refs.~\cite{Cao:2009uw, Goodman:2010qn, Cheung:2010ua, Zheng:2010js, 
Cheung:2011nt, Yu:2011by, Fan:2010gt, Fitzpatrick:2012ix}).

However, the situation is substantially different in LHC searches for DM.
In fact,  effective operators are a tool to describe
the effects of heavy particles (or `mediators') in the low energy theory where these particles
have been integrated out. 
But the LHC machine delivers scattering events at  energies so high, that
they may directly produce the mediator itself. Of course, in this case 
the EFT description fails. This simple point calls for a careful and consistent use of the EFT,
checking its range of validity, in the context
of DM searches at the LHC.

To better illustrate this point,  let us give a simple example, and consider a model with 
a heavy mediator connecting two DM particles to two quarks. The mediator has
mass $\Mmed$ and couplings to quarks and DM $g_q$ and $g_\chi$, respectively. 
At low energies,  much smaller than $\Mmed$, the heavy mediator can be integrated out from the theory 
and one is left with a theory without the mediator, where the interactions between DM and quarks
are described by a tower of effective operators.
The parameters of the ultra-violet (UV) theory including the mediator are connected to the scale $\Lambda$ associated with the dimentions-6 operators of the low-energy EFT through, 
via
\be
\Lambda=\frac{\Mmed}{\sqrt{g_q g_\chi}}.
\label{matching}
\ee
The EFT is valid as long as the events producing DM are such that the mediator cannot be directly
produced. Therefore, they must occur with a momentum transfer $Q_{\rm tr}$
such that
\be
Q_{\rm tr}<\Mmed.
\label{conditionQM}
\ee 
The expansion in terms of a tower of higher-dimensional effective operators can be viewed
as the expansion of the  propagator of the mediator particle,
\be
\label{ki}
\frac{1}{Q_{\rm tr}^2-\Mmed^2}
=-\frac{1}{\Mmed^2}\left(1+\frac{Q_{\rm tr}^2}{\Mmed^2}+\mathcal{O}\left(\frac{Q_{\rm tr}^4}{\Mmed^4}\right)\right)\,.
\ee
Retaining  only the leading  term $1/\Mmed^2$ corresponds to truncating the expansion to the 
lowest-dimensional operator. This truncation is a good approximation only if $Q^2_{\rm tr} \ll \Mmed^2$, which with condition (\ref{matching}) becomes
\be
Q^2_{\rm tr} \ll g_q g_\chi \Lambda^2 \sim \Lambda^2,
\label{conditionQL}
\ee
when the couplings $g_q, \, g_\chi$ are at the natural scale of order 1.
Therefore, one can characterize 
the deficiency of the truncation of the full operator tower to the leading term
by evaluating the expansion parameter $Q_{\rm tr}/\Mmed$.

If  one further assumes  $s$-channel  mediator exchange, then
the kinematics of the process imposes $Q_{\rm tr}>2m_{\rm DM}$, so 
the conditions (\ref{matching}) and (\ref{conditionQM}) 
imply
\be
\Lambda > \frac{Q_{\rm tr}}{\sqrt{g_q g_\chi}} > 2 \frac{m_{\rm DM}}{\sqrt{g_q g_\chi}}\,,
\label{kinconstraint}
\ee
which in the extreme case in which the perturbativity condition on the couplings $g_\chi,g_q<4\pi$ is assumed leads to
\be
\Lambda>\frac{m_{\rm DM}}{2\pi}\,.
\label{mover2pi}
\ee
Addressing quantitatively the question of the validity of the EFT suffers from a dependence 
on the (unknown)  couplings of the UV theory, as shown by 
the  conditions  (\ref{conditionQL}) and (\ref{kinconstraint}).

\begin{figure}[t!]
\centering
\includegraphics[width=0.45\textwidth]{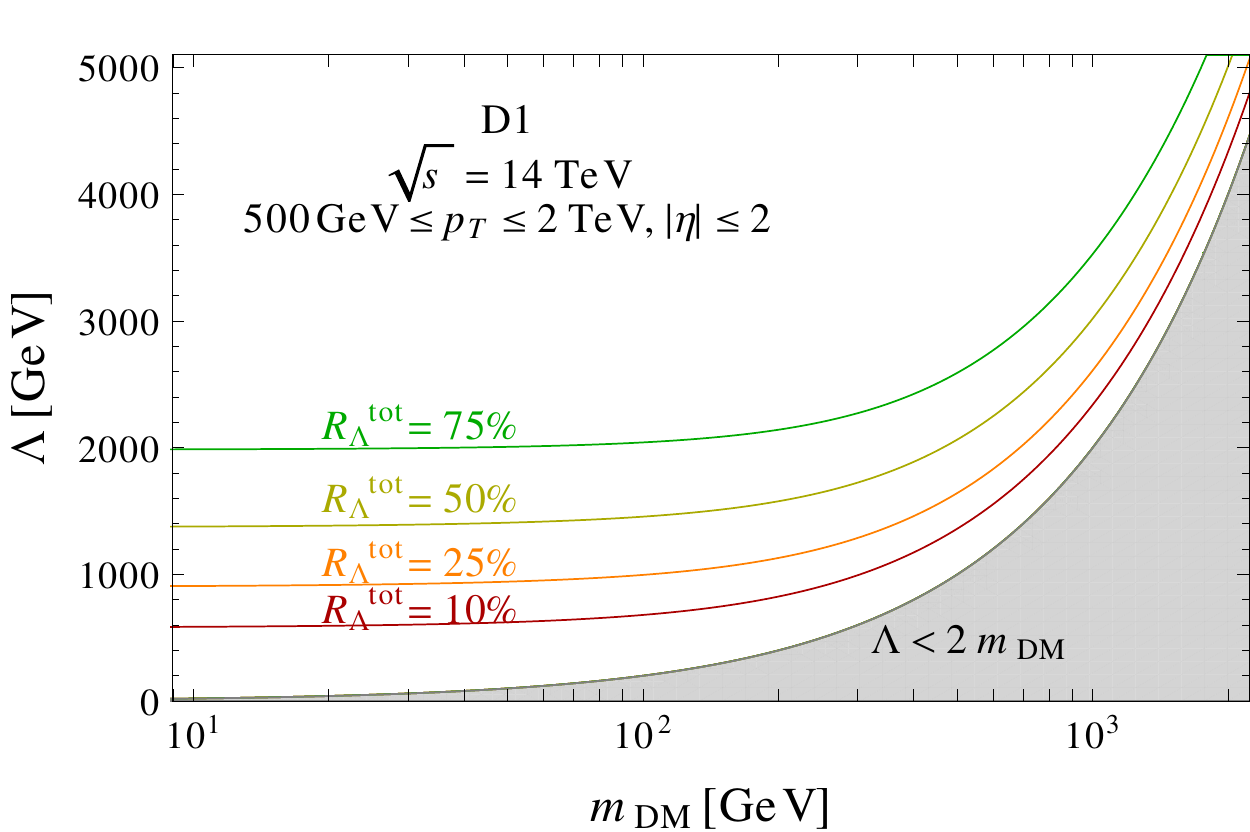} 
\hspace{0.5cm}
\includegraphics[width=0.45\textwidth]{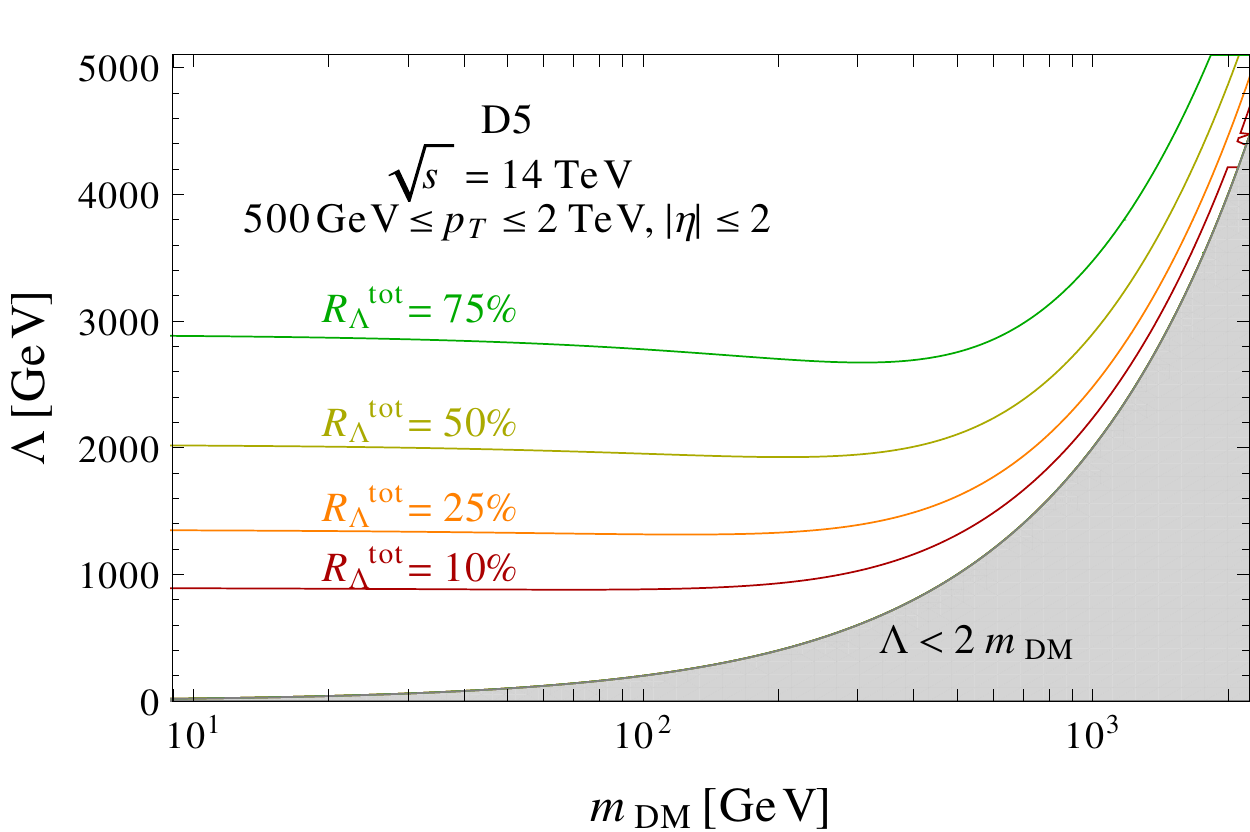} 
\caption{
Contours of the parameter $R_\Lambda^{\rm tot}$, for the D1 (left) and D5 (right) operators and the process $q \bar q \rightarrow \chi \chi  + g$ at center of mass energy $\sqrt{s} = 14$ TeV. Cuts are  chosen to be comparable to those used by the ATLAS and CMS collaborations. }
\label{fig:RLambda}
\end{figure}

With this caveat in mind, it is nonetheless possible to 
quantify the error introduced by using effective operators when describing processes at very high
$Q_{\rm tr}$. 
For example, for a given process, one can calculate the fraction of events that pass the condition (\ref{conditionQL}). We define this as 
\be
R_\Lambda^{\rm tot} \equiv \frac{\sigma|_{Q_{\rm tr} < \Lambda}}{\sigma},
\ee
where $\sigma$ is the cross section for the process of interest, and $\sigma|_{Q_{\rm tr} < \Lambda}$ is the same cross section truncated such that all events pass the condition $Q_{\rm tr} < \Lambda$. 
As an example, contours in $R_\Lambda^{\rm tot}$ are shown in Fig.~\ref{fig:RLambda}, reproduced from ref.~\cite{Busoni:2014sya}, for the process $q \bar q \rightarrow \bar{\chi} \chi + g$, assuming couplings of order unity. The plots correspond to the D1 ($\bar{q}q\bar{\chi}\chi$) and D5 ($\bar{q}\gamma^{\mu}q\bar{\chi}\gamma_{\mu}\chi$) operators in the notation of ref.~\cite{Goodman:2010ku}. The results are qualitatively similar between the different operators and clearly indicate that LHC searches for DM are operating well within the region where the EFT approximation breaks down.

In Refs. \cite{Busoni:2013lha, Busoni:2014sya,Busoni:2014haa,ATL-PHYS-PUB-2014-007} the reader can find an expanded discussion  along these lines, and both analytical and numerical results showing the parameter space regions where
the effective description is valid. It is by now clear that the EFT is not the ideal tool to interpret the LHC data on DM.
A way out of this impasse is  to shift to simplified models, the subject of this paper.
 
%


\section{Simplified Models for $\left(\bar{q}\Gamma^m q\right)~\left(\bar{\chi} \Gamma'_m \chi\right)$ type operators - $s$ - channel model}
\label{sec:qqchichi_s}

\begin{figure}[h!]
  \begin{center}
      \subfigure[]{\scalebox{1.4}{\includegraphics[width=1.7in]{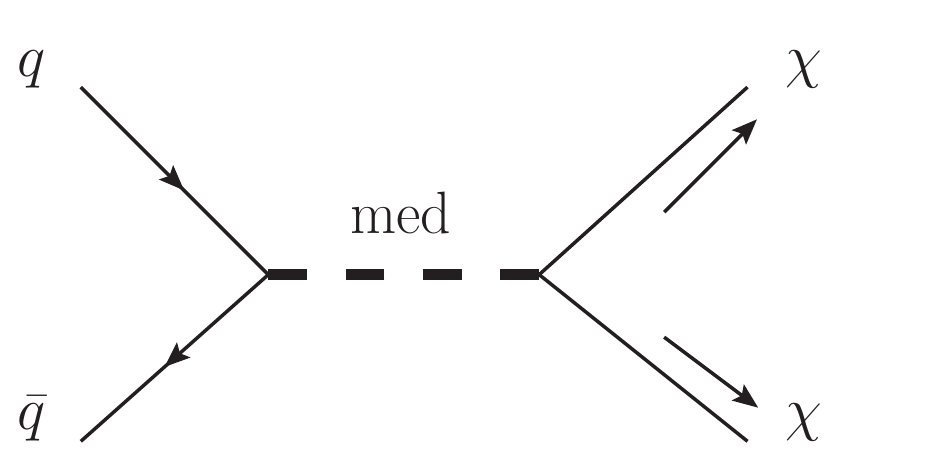}}}
      \subfigure[]{\scalebox{1.4}{\includegraphics[width=1.7in]{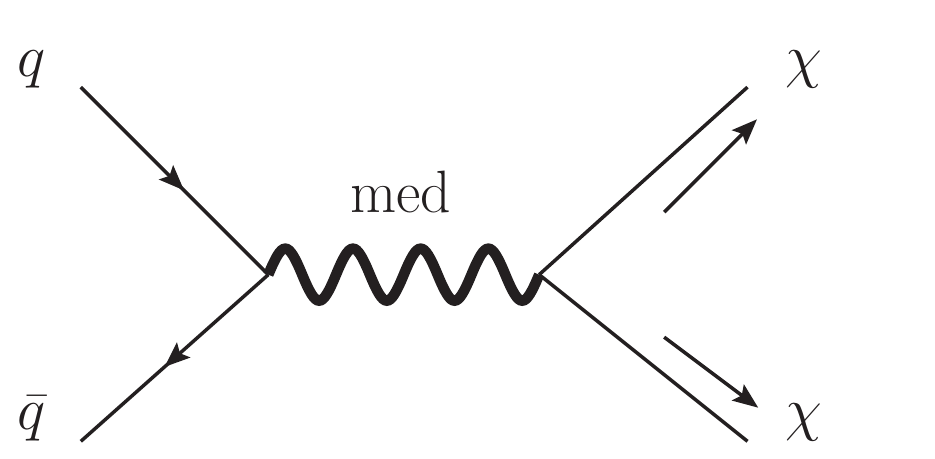}}}
     \caption{Two possible mediators. A massive scalar (left) and/or a massive vector-boson (right), resolving dimension-6 operators of the form, Eq.~(\ref{eqn:dim_six_op}), through an $s$-channel exchange. }
\label{fig:simp_mod_s_channel}
\end{center}
\end{figure}

Modeling the DM particle, $\chi$, as a fermion we consider the dimension six operators of the form,
\be
\label{eqn:dim_six_op_s-ch}
\mathcal{O}_6 = \left(\bar{q}\Gamma^m q\right)~\left(\bar{\chi} \Gamma'_m \chi\right) \, .
\ee
These operators are the D1-D10 (and D1$^\prime$-D4$^\prime$) operators in the notation of Refs.~\cite{Goodman:2010ku, Busoni:2014sya}. The simplest way of resolving four-fermion operators as in Eq.~(\ref{eqn:dim_six_op_s-ch}) is through a color-singlet boson, either a scalar or a vector, as shown in Fig.~\ref{fig:simp_mod_s_channel}. The simplified model we describe assumes CP-conservation and contains a new scalar (pseudoscalar), $S$ ($S'$), or a new vector (axial-vector), $V_\mu$ ($V_\mu^\prime$), with interactions,
\be
\label{eq:Ssimp}
\mathcal{L}_{\rm S}		& \supset &	-\frac{1}{2}M_{\rm med}^2S^2-y_{\chi}S\bar{\chi}\chi-y_q^{ij}S\bar{q}_iq_j+{\rm h.c.}\,, \nonumber \\
\mathcal{L}_{\rm S'}	& \supset &	-\frac{1}{2}M_{\rm med}^2S^{\prime2}-y^\prime_\chi S^\prime\bar\chi\gamma_5\chi-y_q^{\prime ij}S\bar{q}_i\gamma_5q_j+{\rm h.c.}\,, \nonumber \\
\mathcal{L}_{\rm V}		& \supset &	\frac{1}{2}M_{\rm med}^2V_\mu V^\mu-g_\chi V_\mu\bar\chi\gamma^\mu\chi-g_q^{ij}V_\mu\bar q_i\gamma^\mu q_j\,, \nonumber \\
\mathcal{L}_{\rm V'}	& \supset &	\frac{1}{2}M_{\rm med}^2V^\prime_\mu V^{\prime\mu}-g^{\prime}_\chi V^\prime_\mu\bar\chi\gamma^\mu\gamma_5\chi-g_q^{\prime ij}V^\prime_\mu\bar q_i\gamma^\mu\gamma_5q_j\,.
\ee
where $q=u,d$ and $i,j=1,2,3$ are flavor indices. Such simplified models have been considered in several past publications, see for example the early work of ref.~\cite{Bai:2010hh,Fox:2011pm,Goodman:2011jq} as well as more recent works~\cite{Abdullah:2014lla, Buchmueller:2014yoa} and references therein. These Lagrangian terms generate the effective operators D1$^\prime$, D4$^\prime$, D5 and D8. Refs.~\cite{Busoni:2013lha,Busoni:2014sya} find that the operators (D2$^\prime$, D3$^\prime$) and (D6, D7) have the same partonic level cross section as (D4$^\prime$, D1$^\prime$) and (D8, D5), respectively. We thus do not include the former in what follows. Note that a UV complete description of scalar theory would require $y_q\simeq m_q/M_{\rm med}$ (resulting in the operators D1-D4), but since the translation between these cases is simple, we find the use of Eqs.~\eqref{eq:Ssimp} sufficient for our purposes.

As concerns the mediator couplings to quarks, the existence of off-diagonal coupling is tightly constrained by various FCNC processes~\cite{Isidori:2010kg}. We do not study such couplings, taking $g_q^{ij}=g_q^i\delta^{ij}$. In the following we consider the scenario of flavor blind couplings to all quarks: $g_d^i=g_u^i\equiv g_q$ for $i=1,2,3$. An interesting scenario, in which the mediator couples more strongly to the third generation is discussed below in Sec.~\ref{sec:third_generation}. We further assume that the only available decay channels of the mediator are into quarks and DM particles.

The differential cross sections at the parton level (with respect to the pseudo-rapidity ($\eta$) and transverse momentum ($p_{\rm T}$) of the final jet) for the $s$-channel process $f(p_1)+\bar{f}(p_2)\rightarrow\chi(p_3)+\chi(p_4)+g(k)$ are given in Eqs.~(2.4)-(2.8) of ref.~\cite{Busoni:2014sya}, where
\be
\Lambda^4	& = &	\frac{\left(Q_{\rm tr}^2-M_{\rm med}^2\right)^2+\Gamma^2M_{\rm med}^2}{g_q^2g_\chi^2}\, ,
\ee
should be used to resolve the EFT operators. 

As discussed earlier, the EFT approach, where integrating out a heavy mediator generates a tower of higher dimensional operators, is appropriate in processes with low energy transfer:
\be
M_{\rm med} \gtrsim Q_{\rm tr} \geq 2 m_\chi  \,.
\ee
Refs.~\cite{Busoni:2013lha,Buchmueller:2013dya, Busoni:2014sya} discuss the limitations of the EFT approach for DM searches for an $s$-channel mediator exchange, and quantify the dependence of the errors resulting from the EFT approach on mediator and DM masses and couplings. At the partonic level the differences between the cross sections of the effective theory and the full theory are,
\be
\left(\frac{\d^2\hat\sigma}{{\d}\eta {\d} p_{\rm T}}\right)_{\rm full}\Big/\left(\frac{\d^2\hat\sigma}{{\d}\eta{\d} p_{\rm T}}\right)_{\rm EFT}=\frac{M_{\rm med}^4}{\left(Q_{\rm tr}^2-M_{\rm med}^2\right)^2+\Gamma^2M_{\rm med}^2}\, ,
\ee
where $\Lambda=\Mmed/\sqrt{g_q g_\chi}$ was used.

The authors of ref.~\cite{Busoni:2014sya} study the ratio between the EFT resulting cross section and the full theory at 8~TeV center of mass energy. They find that this ratio is smaller by $50\%$ for both scalar and vector interactions if $\Lambda\gtrsim2-3$~TeV and $m_\chi\lesssim1$~TeV. In the following we explore the validity of the EFT approach as a function of the final jet $p_T$ at $\sqrt{s}=14$~TeV. For this high energy, the gluon initiated process is significant and contributes comparably to the quark initiated process, for high $p_T$ cuts. We therefore present numeric results based on Monte Carlo simulated events. The events are generated using {\tt MadGraph~5}~\cite{Alwall:2011uj} imposing a cut of $p_T\geq200$~GeV and $|\eta|\leq2.5$ on the final jet. To quantify the differences between the EFT and the simplified model approaches we use the ratio of partonic level cross-section with a single final state jet in addition to the DM pair. We expect next-to-leading order corrections, showering, hadronization, and detector effects to largely cancel in the ratio, and leave a more detailed analysis to future study.

\begin{figure}[h!]
  \begin{center}
      \subfigure[]{\scalebox{1.8}{\includegraphics[width=2.5in]{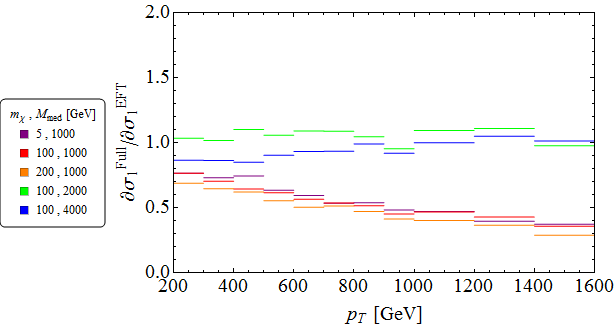}}}
      \subfigure[]{\scalebox{1.8}{\includegraphics[width=2.5in]{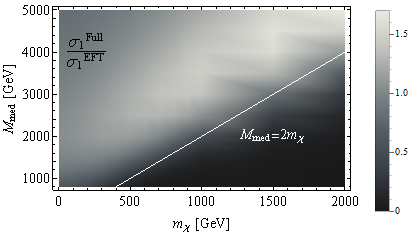}}}
 \caption{The ratio between the interaction cross section of the full theory and the EFT one in the case of scalar mediator. The top figure shows the ratio of differential cross sections, integrated over the jet rapidity $-2.5\leq\eta\leq2.5$. The ratio of the total cross section integrated over the jet transverse momentum $p_{\rm T}\geq200$~GeV is plotted in the bottom figure. The events are generated using {\tt MadGraph~5}.}
 \label{fig:FullvsEFT1}
  \end{center}
\end{figure}

\begin{figure}[h!]
  \begin{center}
      \subfigure[]{\scalebox{1.8}{\includegraphics[width=2.5in]{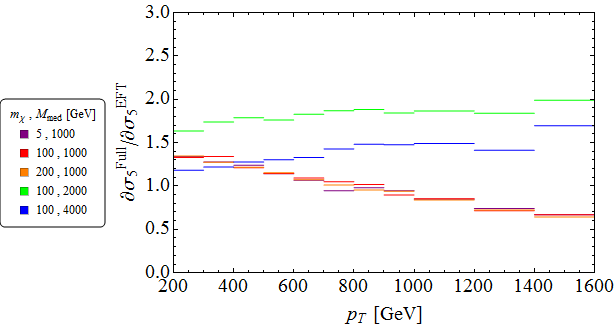}}}
      \subfigure[]{\scalebox{1.8}{\includegraphics[width=2.5in]{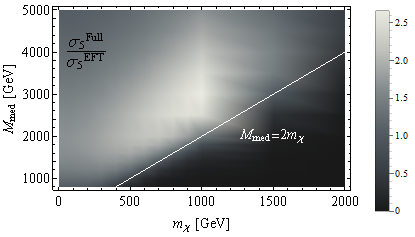}}}
\caption{The same as Fig~\ref{fig:FullvsEFT1} but for a vector mediator.}
\label{fig:FullvsEFT5}
  \end{center}
\end{figure}

Fig.~\ref{fig:FullvsEFT1}~(\ref{fig:FullvsEFT5}) shows the ratio between the interaction cross sections resulting from the simplified model and the effective theory for the scalar (vector) mediated interactions. At the top pane we present this ratio of the differential cross sections as a function of the jet $p_{\rm T}$, for several choices of DM and mediator masses. At the bottom pane we show the ratio between the total cross section as a function of the DM and mediator masses. It can be seen that the two approaches coincide for $M_{\rm med}\gg2m_\chi,p_{\rm T}$. However, if this condition is not fulfilled, differences between the full theory and the EFT approach appear both in the total cross section and in the kinematical distribution of the two. It is thus necessary to go beyond the EFT study in order to correctly explore the region of parameter space where $M_{\rm med}\lesssim2m_\chi$.

To find the most convenient and enlightening set of simplified models, one needs to study the sensitivity of the observables to the helicity structure of the mediator couplings. For the scalar and pseudoscalar interactions, Refs.~\cite{Busoni:2013lha,Busoni:2014sya} find, at the parton level,
\be
\left(\frac{\d^2\hat\sigma}{{\d}\eta{\d} p_{\rm T}}\right)_{D_1^\prime}	/ \left(\frac{\d^2\hat\sigma}{{\d}\eta{\d} p_{\rm T}}\right)_{D_4^\prime} & = &	 \left( 1- \frac{4m_{\rm DM}^2}{Q_{\rm tr}^2} \right)  \, ,
\ee
while for vector couplings
\be
\left(\frac{\d^2\hat\sigma}{{\d}\eta{\d} p_{\rm T}}\right)_{D_5} / \left(\frac{\d^2\hat\sigma}{{\d}\eta{\d} p_{\rm T}}\right)_{D_8} & = &	\left( \frac{Q_{\rm tr}^2 + 2m_{\rm DM}^2}{Q_{\rm tr}^2 -4m_{\rm DM}^2} \right)  \, .
\ee
In the limit that $Q_{\rm tr}\gg2m_{\rm DM}$, the two differential cross sections share the same $\eta$ and $p_{\rm T}$ distribution. However, these kinematical regions are suppressed by the parton distribution functions (PDFs). 

\begin{figure}[h!]
  \begin{center}
      \subfigure[]{\scalebox{1.8}{\includegraphics[width=2.5in]{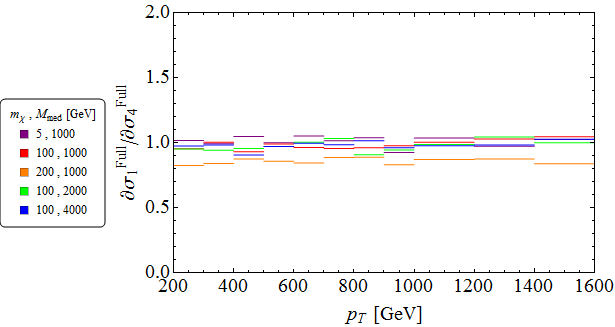}}}
      \subfigure[]{\scalebox{1.8}{\includegraphics[width=2.5in]{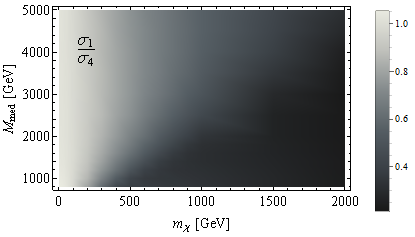}}}
\caption{The ratio between the $s$ channel interaction cross section mediated by a scalar and a pseudoscalar mediator. The top figure shows the ratio of differential cross sections, integrated over the jet rapidity $-2.5\leq\eta\leq2.5$. The ratio of total cross sections integrated over the jet transverse momentum $p_T\geq100$~GeV is plotted in the bottom figure.}
\label{fig:1vs4}
  \end{center}
\end{figure}

\begin{figure}[h!]
  \begin{center}
\subfigure{\scalebox{1.8}{\includegraphics[width=2.5in]{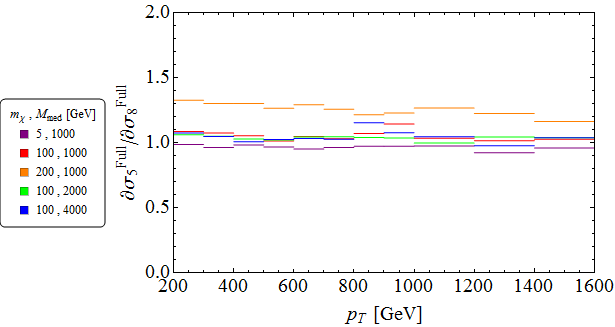}}}
\subfigure{\scalebox{1.8}{\includegraphics[width=2.5in]{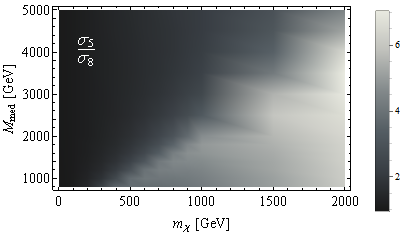}}}
\caption{The same as Fig~\ref{fig:1vs4} but for a vector and an axial vector interactions.}
\label{fig:5vs8}
\end{center}
\end{figure}

To explore the impact of the different helicity structures we study the ratio between the cross sections arising from scalar (vector) and pseudoscalar (axial vector) mediators. The results, based on events generated using {\tt MadGraph~5}, are shown in Figs.~\ref{fig:1vs4} and~\ref{fig:5vs8} for the scalar and vector cases, respectively. As above, we present this ratio for the differential cross section as a function of the jet $p_{\rm T}$ at the top, and as a function of the DM and mediator masses at the bottom. As expected, we find that the ratios between each pair of cross sections ({\it i.e.} scalar vs. pseudoscalar, and vector vs. axial vector) have only weak dependence on the final jet $p_{\rm T}$. The processes do however have different overall cross sections and thus will result in different number of signal events. Furthermore, this ratio has a nontrivial dependence on $M_{\rm med}$ for heavy DM particles, as a result of the PDFs. Since the scalar and axial vector interactions result in smaller cross-sections it is sufficient, as a first step, to explore the scalar and axial vector mediation resolving the $s$-channel DM pair-production at the LHC. If a signal is discovered, further analysis of the jet angular distribution could differentiate between the different particles mediating the DM production.

Simplified models with $s$-channel mediator might also leave significant footprints in various experimental searches other than direct DM production at colliders. These are, for example, direct DM detection experiments and resonance searches in dijet production at the LHC. While it is important to consider these additional searches, this should be done with care since the reinterpretation of a set of constraints is model dependent. For instance, direct detection constraints are significantly weakened if the interaction is spin-dependent. Furthermore, since the collision energies are much lower ($1-100$ keV range), direct detection may be entirely evaded if the dark spectrum is split by more than $100$~keV or so, as is the case in inelastic DM models~\cite{TuckerSmith:2001hy}. On the other hand such mass splittings are not a barrier at colliders and models of this kind can be searched for at the LHC (see further discussion in section 4.A of ref.~\cite{Weiner:2012cb}). This therefore provides another cogent example of the importance of a comprehensive program of complimentary DM searches. Sec.~\ref{sec:mediators} and references therein contain a detailed discussions of the searches for the mediators and various experimental constraints. Here we briefly describe the constraints coming from dijet searches at the LHC.

The dijet narrow resonance searches, \cite{Aad:2011fq,Aad:2014aqa}, are relevant only for $\Gamma_{\rm med}/M_{\rm med}\lesssim 0.15$ in case of Gaussian shape or $\Gamma_{\rm med}/M_{\rm med}\lesssim0.05$ for the Breit-Wigner case. In other cases the mediator will escape the direct searches. Therefore, the maximal couplings that may be probed by the narrow resonance searches with a Gaussian shape are
\be
	y_q < 1.1/\sqrt{N_q} \, , \qquad g_q < 1.4/\sqrt{N_q} \, .
\ee
For the Breit-Wigner case we find, 
\be
	y_q < 0.65/\sqrt{N_q} \, , \qquad g_q < 0.79/\sqrt{N_q} \,.
\ee
Here we assume that the couplings to $N_q$ quarks are equal to $y_q(g_q)$ for scalar and pseudoscalar (vector/axial vector) and both phase-space effects and the coupling to the DM candidate are neglected.

The CMS dijet angular distribution~\cite{Chatrchyan:2012bf} and the ratio between the central and forward dijet cross-sections given by ATLAS~\cite{Aad:2011aj} can be used to constrain models which can escape the narrow width searches. We note that the sensitivity of the angular distribution for relatively light mediators, $M_{\rm med}\lesssim1\,$TeV, is limited because of the large contribution from gluon fusion. In that case, it may be that Tevatron data  can be used to better constrain the relevant parameter space, as discussed in ref.~\cite{Dobrescu:2013cmh} for example.

\subsection{Expected Sensitivity for Monojet Search at 14 TeV}
\label{sec:bounds}
\begin{figure}[t!]
\centering
\includegraphics[width=0.6\textwidth]{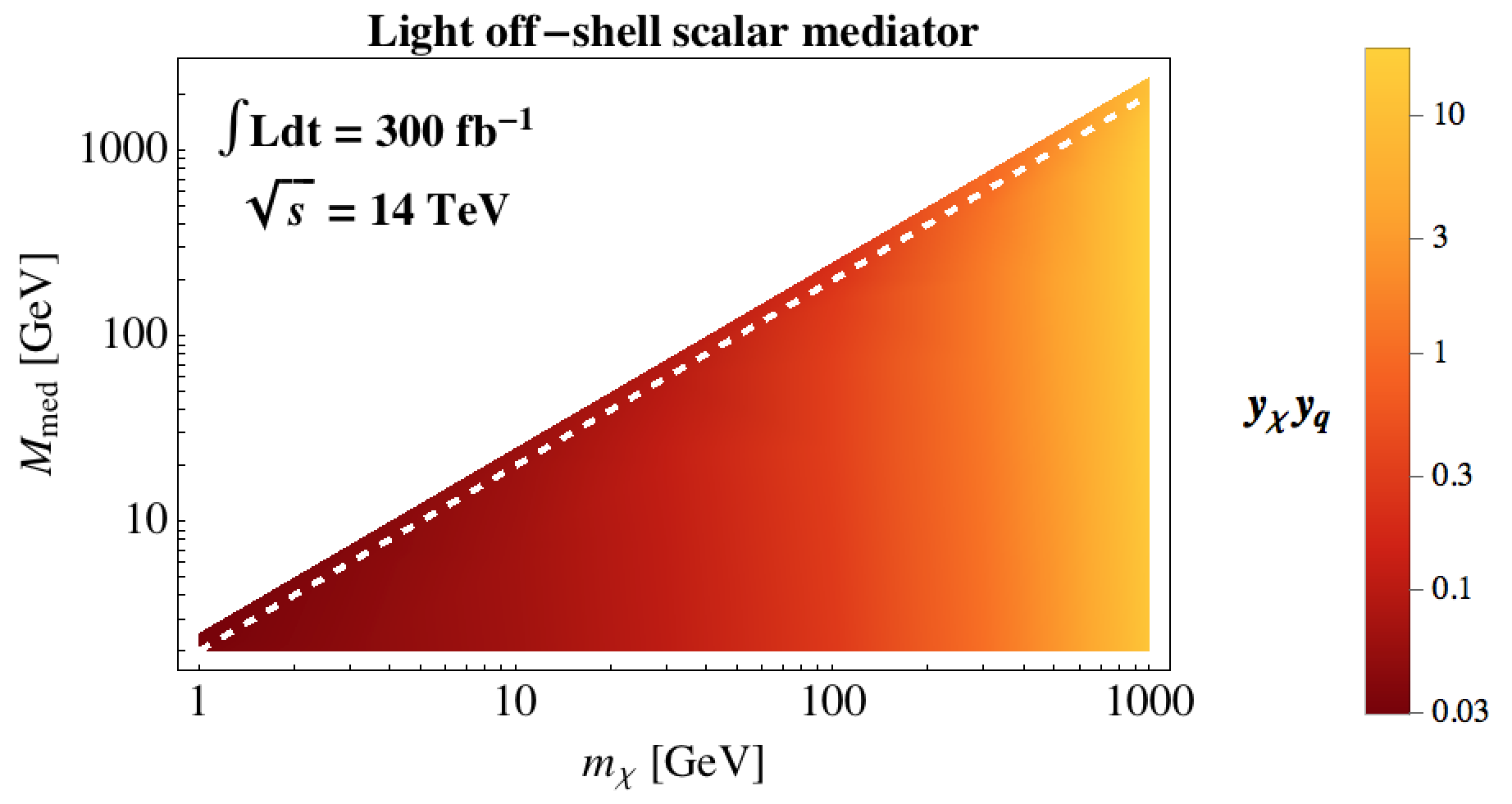} 
\hspace{0.5cm}
\includegraphics[width=0.6\textwidth]{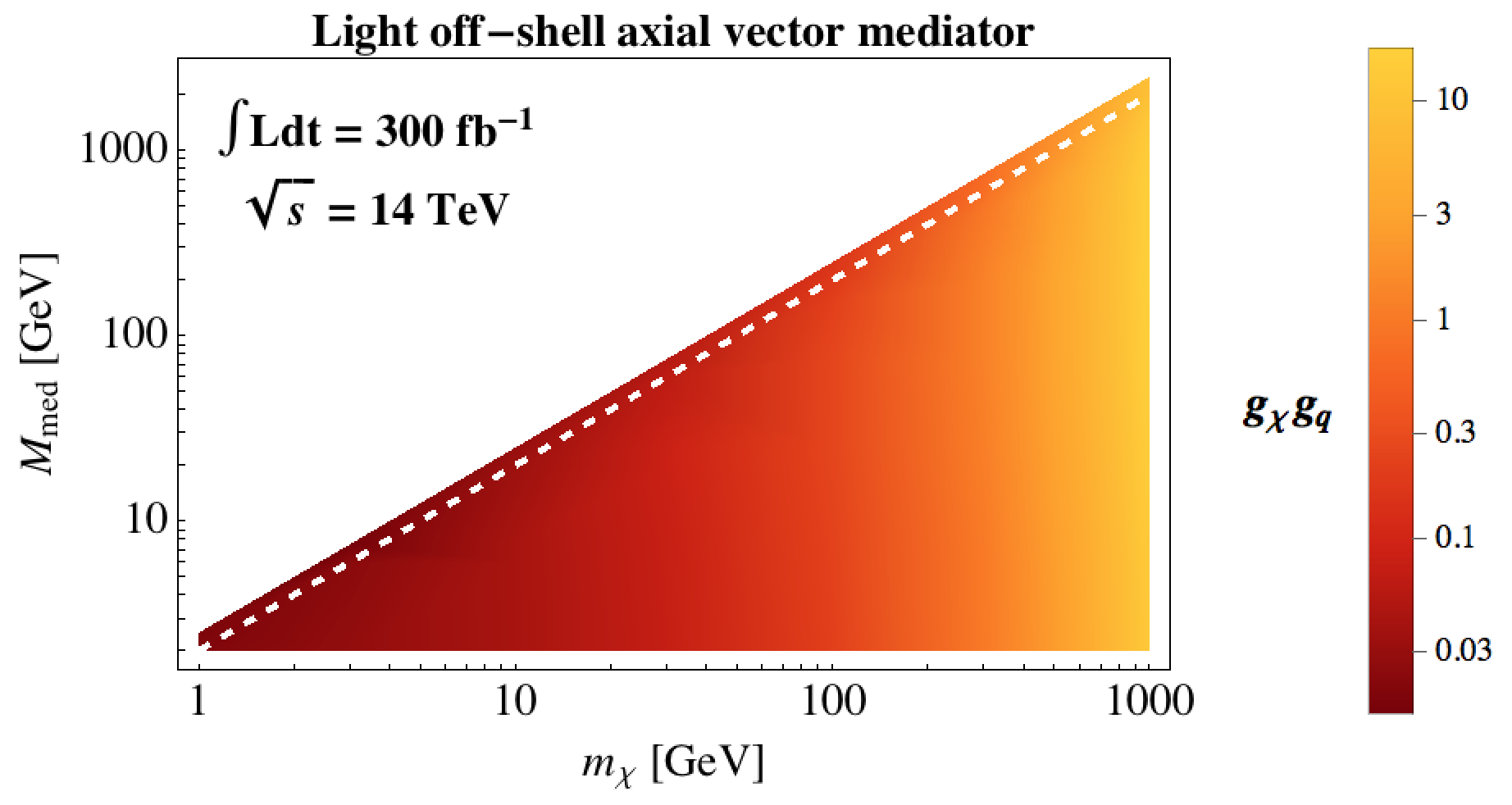} 
\caption{
Expected sensitivity at $95\%$ CL to the product of mediator coupling to DM and SM in the off-shell regime.  The limits are derived for run II of the LHC at $14 \text{ TeV}$ with $300 \text{ fb}^{-1}$ of data, assuming a scalar (top) and an axial vector (bottom) mediator.  The white dashed line indicates the boundary where $M_{\rm{med}} = 2m_\chi$.
}
\label{fig:monojet_offshell_bounds}
\end{figure}

We present the expected sensitivity for a DM search in events with a monojet and MET ($\slashed{E}_T$)  for the $\sqrt{s} = 14$ TeV run of the LHC and integrated luminosity of $300\text{ fb}^{-1}$. The limits, shown in Fig.~\ref{fig:monojet_offshell_bounds}, are presented using simplified models with a scalar or axial vector mediator. Using Monte Carlo simulations for both background and signal, we derive prospective limits at $95\%$ confidence-level (CL) on the product of mediator to DM and mediator to SM couplings, $y_\chi y_q$ ($g_\chi g_q)$ for scalar (axial vector) mediator, for a range of DM and mediator masses. We assume flavor independent couplings to quarks and consider only part of the parameter space where predominantly off-shell production of DM occurs, i.e. $M_{\rm{med}} \lesssim 2 m_\chi$. In this regime the cross section is expected to be independent of the mediator width, except for the region where $M_{\rm{med}} \approx 2m_\chi$. In order to guarantee pertubativity of the models we only probe the parameter space for which,
\begin{equation}
g_{\chi/q}, y_{\chi/q} \leq 4\pi
\end{equation}
and
\begin{equation}
\frac{\Gamma_{\rm{med}}}{M_{\rm{med}}} \lesssim 0.5.
\label{eq:gamma}
\end{equation}
In the region where the production cross section depends on $\Gamma_{\rm{med}}/M_{\rm{med}}$, we take the pertubative limit, $\Gamma_{\rm{med}}/M_{\rm{med}}=0.5$. In the heavy DM mass region ($m_\chi \gtrsim 800 \text{ GeV}$), where $y_\chi y_q \gtrsim 10.5$ for the scalar mediator and $g_\chi g_q \gtrsim 13$ for the axial vector mediator (which is outside our parameter space), one finds no sensitivity in the pertubative regime unless the mediator couples only to light quarks thereby suppressing $\Gamma_{\rm{med}}/M_{\rm{med}}$.

The main background processes are $Z \rightarrow \nu \bar{\nu} + jets$, $W \rightarrow \ell\nu + jets$, where $\ell=e,\mu,\tau$, and single boson production ($Z, W$) with the jet coming from Initial State Radiation (ISR). We consider these in the leading order approximation.  Other background processes, such as di-boson and $t\bar{t}$~+~single~top were not taken into consideration, as their contribution to the background is smaller by orders of magnitude~\cite{ATLAS:2012zim}. 

The background and signal events were generated using {\tt MadGraph~5} generator~\cite{Alwall:2011uj} (with MSTW2008 PDF) for the hard process, and {\tt{Pythia}}~6~\cite{Sjostrand:2006za} for showering and hadronization. For both signal and background we match the one and two jet samples. After the event generation, the interaction of the generated particles with the detector material and the detector response were simulated with {\tt{Delphes}}~3.1.2~\cite{Ovyn:2009tx} and {\tt{ROOT}}~5.3.4~\cite{Antcheva:2009zz}, customized to the ATLAS detector geometry. All events were required to have $\slashed{E}_T  > 120$~GeV, at least one jet with $p_{\rm{T}} > 130$~GeV and $|\eta|<2.0\,$. Events with more than two jets and events with a muon or an electron with $p_{\rm{T}}>30\,$GeV and $|\eta|<2.0$ were rejected. For signal and background events, the leading jet $p_{\rm{T}}$ distribution was drawn using the same binning as in the ATLAS monojet analysis~\cite{ATLAS:2012zim}. The limit on the product of the coupling constants, $y_\chi y_q$ and $g_\chi g_q$, was calculated by requiring that the probability to find the background plus signal $p_{\rm T}$ distribution assuming the background only hypothesis gives a $p$-value of 0.05 using Poisson statistics. We leave a more detailed analysis including the case of on-shell production to a future study. 

\section{Simplified Models for $\left(\bar{q}\Gamma^m q\right)~\left(\bar{\chi} \Gamma'_m \chi\right)$ type operators - $t$ - channel models}
\label{sec:qqchichi_t}

\begin{figure}[t]
\begin{center}
\includegraphics[width=0.4 \textwidth]{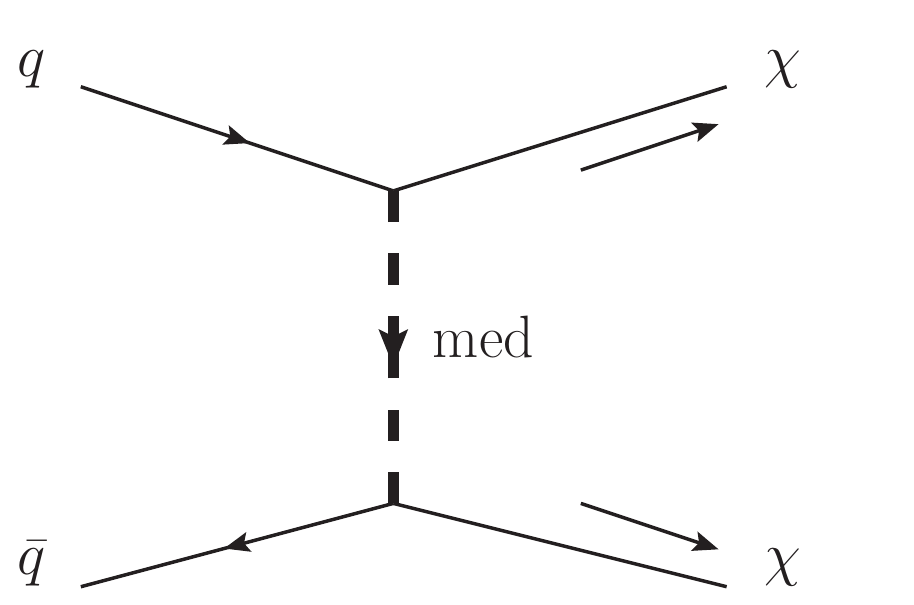}
\end{center}
\caption{A massive scalar resolving dimension-6 operators of the form, Eq.~(\ref{eqn:dim_six_op}), through a $t$-channel exchange. The MSSM with its squarks and neutralinos is an example of a full model exhibiting such a process. }
\label{fig:simp_mod_t_channel}
\end{figure}

Modeling the WIMP $\chi$ as a fermion, we consider the dimension-6 operators of the form,  
\be
\label{eqn:dim_six_op}
\mathcal{O}_6 = \left(\bar{q}\Gamma^m q\right)~\left(\bar{\chi} \Gamma'_m \chi\right)
\ee
These are the D1-D10 operators in the notation of ref.~\cite{Goodman:2010ku}. These effective operators are generated through some new dynamics such as a particle mediating the interaction at tree level. In this section we consider a colored fermionic mediator with an interaction vertex between quarks and the WIMP resulting in a $t$-channel exchange as shown in Fig.~\ref{fig:simp_mod_t_channel}. Similar to the $s$-channel case, this process can be searched for in events with large missing energy~\footnote{One important difference between $s$- and $t$-channel mediators is that in the latter case colored radiation can originate from the mediator itself. See ref.~\cite{Papucci:2014iwa} for the full set of leading order diagrams contributing to the process.}. A concrete model is that of a squark exchange in supersymmetric models:
\be
\mathcal L = \mathcal{L}_{SM} + g_{M}\sum_{i=1,2} \left(\widetilde{Q}_L^i \bar{Q}^i_L +  \tilde{u}^i_R \bar{u}^i_R + \tilde{d}^i_R \bar{d}^i_R \right)\chi + \text{mass terms} +c.c. 
\label{eq:lagrangian}
\ee 
where $Q_{L}^i, u_R^i, d_R^i$ are the usual SM quarks, $\widetilde{Q}_L^i, \widetilde{u}_R^i, \widetilde{d}_R^i $ correspond to the respective squarks (from hereon the ``mediators"), and $i$ represents an index running over the first two generations, since we will not look at signals involving the third generation (see Sec.~\ref{sec:third_generation}). Unlike the usual case in Superysmmetry, here the WIMP $\chi$ can be taken to be either Dirac or Majorana fermion.   For simplicity we will take the mediator masses to be degenerate and focus on two different extreme cases: 1) all mediator flavors are present or 2)  only $\tilde d_{R}^{i}$ are present. Simply due to multiplicity these two cases maximize and minimize the mediator production cross-section, respectively. 

Since it is coupled to quarks, if kinematically possible the mediator can be produced on resonance at colliders. As discussed in section~\ref{sec:problems_with_EFT}, in this regime the EFT is no longer valid and the question is whether the actual limits are substantially changed. The largest production cross-section is associated with regions that are both at low center of mass energy (due to PDF effects), and on resonance~\cite{Petriello:2008pu,Gershtein:2008bf, Dreiner:2013vla, Busoni:2013lha, Buchmueller:2013dya, Chang:2013oia, An:2013xka, Bai:2013iqa, DiFranzo:2013vra}. On the other hand, signal events at low $\sqrt s$ are strongly contaminated by SM backgrounds. As discussed in ref.~\cite{Papucci:2014iwa}, by comparing the EFT to the simplified model, one generally finds that the constraints from the simplified model are markedly different compared with those extracted from the EFT.  If the mediator is kinematically accessible but sufficiently heavy, the correct bounds are stronger than those extracted from the EFT on account of resonant production~\cite{Fox:2011pm}.  If the mediator is light then the signal appears in the region contaminated by background and the EFT constraints are overly strong~\cite{An:2012va}. 

As in the case of $s$-channel discussed in Sec.~(\ref{sec:qqchichi_s}) when the mediator is kinematically accessible one can directly search for it in other final states. Such searches may have a stronger impact than the monojet searches.  For example, since the mediator couples to quarks and/or is a colored particle, this means that the mediator, rather than decaying only to DM, may be pair produced and be detected in multi-jet events with large missing energy~\cite{Papucci:2014iwa}.   As found in ref.~\cite{Papucci:2014iwa}, while monojet constraints on DD are relatively model independent in the EFT regime (which is not entered until the mediator is above 3 TeV in the $s$-channel case, and 1 TeV in the $t$-channel case and the DM is parametrically lighter), they rarely represent the true constraints, being either too weak (heavier mediator) or too strong (lighter mediator).    \\

The above Lagrangian, Eq.~(\ref{eq:lagrangian}), induces a minimal decay width for each mediator flavor given by the expression
\be
\Gamma_{\rm med}^{min} = \frac{g_M^2 \Mmed}{16 \pi} \left(1-\frac{m_{DM}^2}{\Mmed^2}\right)^2,
\label{Gammamin}
\ee
where $\Mmed$ is the mediator mass. The mediator width can be larger if additional states to which it can decay exist. These additional states are possibly constrained by LHC searches other than the ones considered herein. Since this involves more model dependency, we leave the mediator decay width as a free parameter in our results.

We now briefly review how multi-jet plus missing energy searches can probe the parameter space of this simplified model. Monojet analyses are cut-and-count based and involve signal regions defined by cuts on the transverse momentum of the jet and missing energy in the event. Limits are set independently for each signal region and the upper bound on the number of signal events is provided so that no further statistical analysis is necessary. By simulating the signal with different values of the coupling, $g_M$,  one can find the maximal allowed couplings compatible with observations. We note that as the coupling increases, the width of the mediators must be taken at least as large as $\Gamma_{\rm med}^\text{min}$, according to Eq.~(\ref{Gammamin}).  

An important caveat to consider when performing a monojet analysis is that, despite the name, starting from analyses for the 8 TeV run of the LHC, no cuts on the $p_T$ of the second leading jet are imposed. Simulating event samples without a second hard jet at parton level is therefore erroneous and would produce dramatically weaker constraints, as shown in ref.~\cite{Papucci:2014iwa}.

A second remark that applies both to monojet and multi-jet plus MET searches concerns the effect of the narrow width approximation (NWA). The standard procedure taken by ATLAS to extract limits on simplified models (for instance in the searches for gluinos or squarks) is to generate events for on-shell production of the heavier resonance, which later decays into the dark matter plus jets  and missing energy. In doing so, one implicitly assumes that the cross-section is dominated by diagrams with mostly on-shell squarks and that their width is extremely narrow.  On the other hand, the values of the coupling to which jets+MET searches are sensitive to force the squark widths to be comparable or larger than the $p_{T}$ thresholds of the jets required by the analyses. Thus, finite width effects are important and once again we refer to ref.~\cite{Papucci:2014iwa} for a quantitative discussion of these effects.

\begin{figure}[t]
\begin{center}
\includegraphics[scale=0.535]{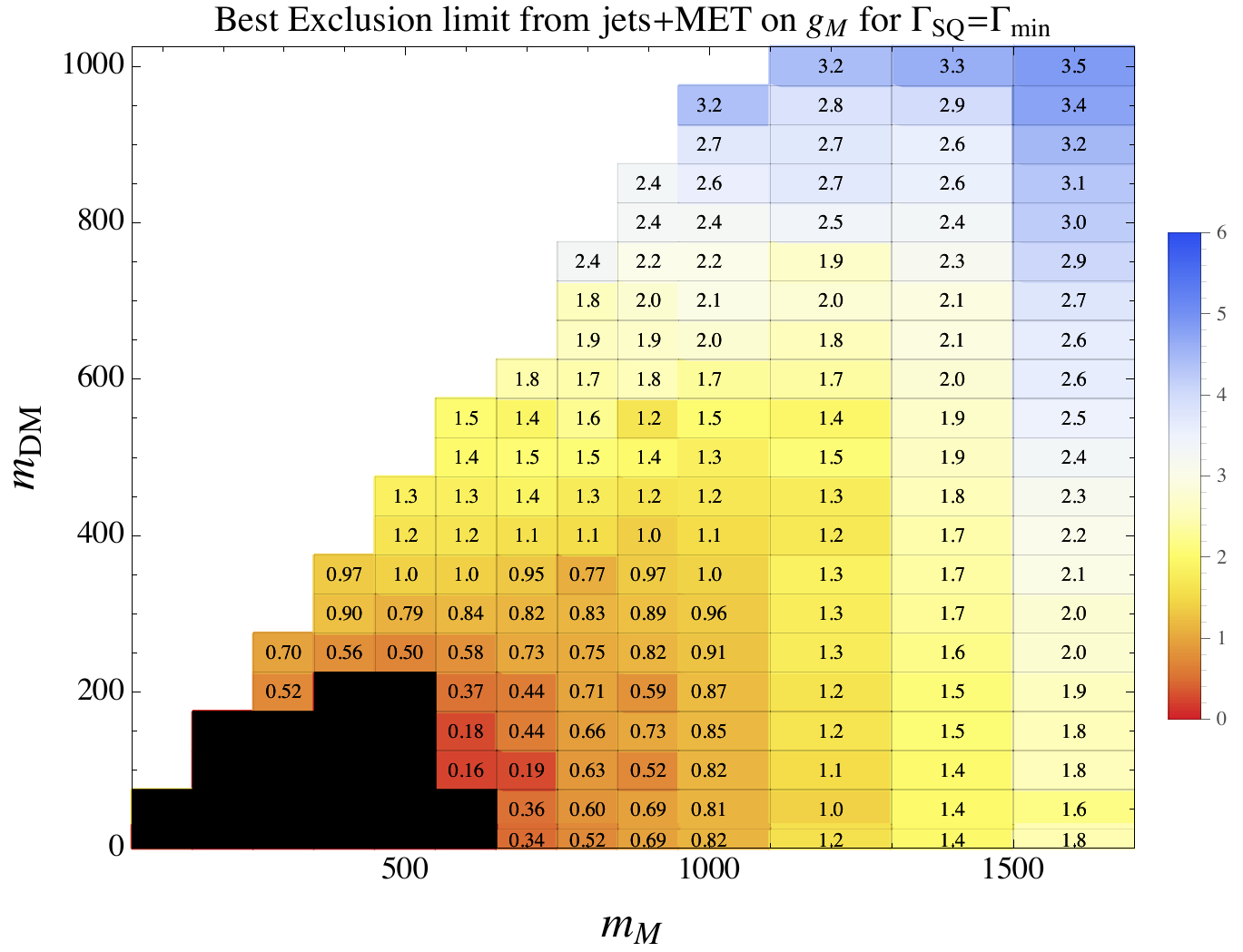}
\includegraphics[scale=0.535]{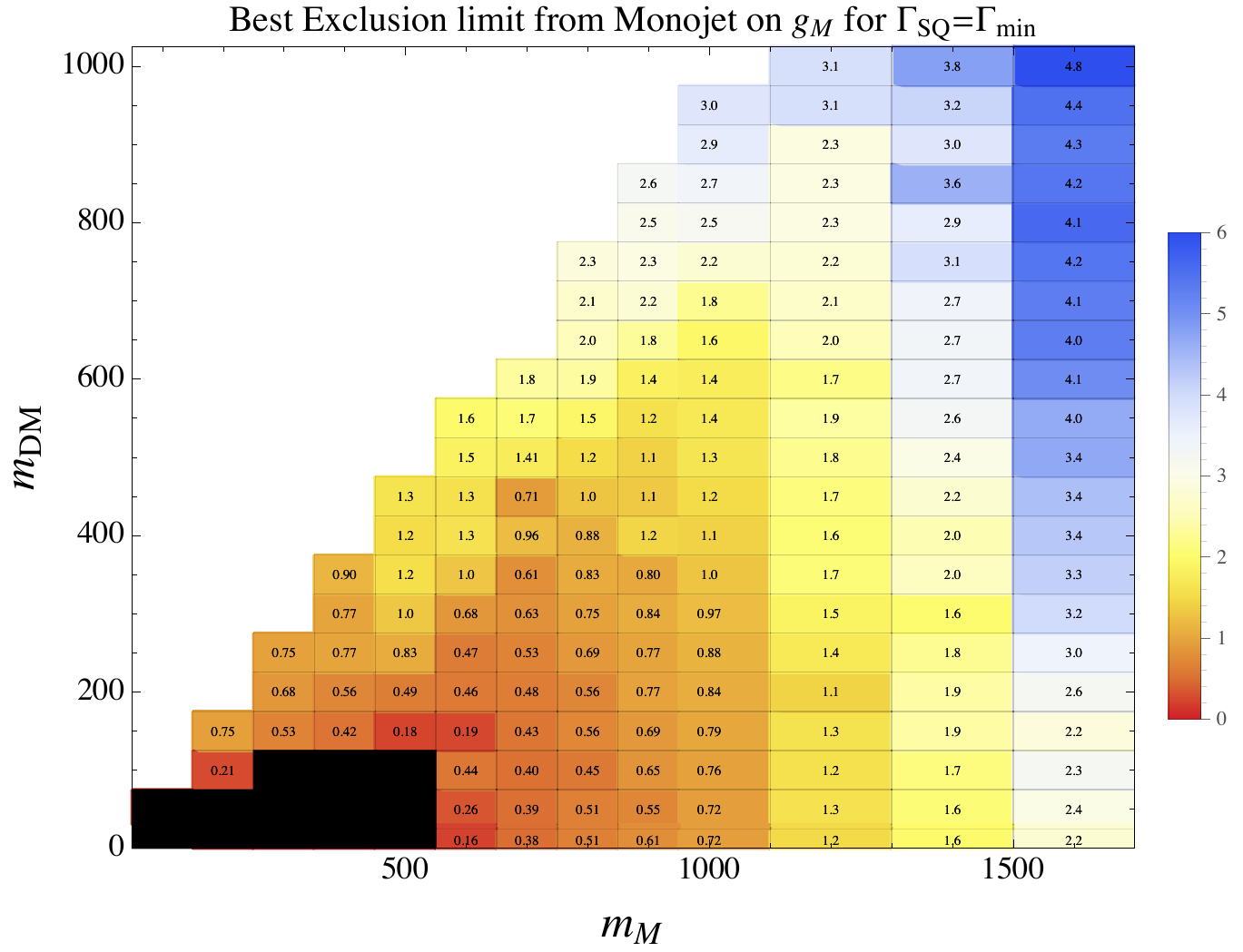}
\end{center}
\caption{Limits from ref.~\cite{Papucci:2014iwa} on $g_{M}$ (for the case of mediator coupling to $\tilde{u},\, \tilde{d},\,\tilde{c},\,\tilde{s}$, $L+R$) from ({\em left}) jets+MET, and ({\em right}) monojet, for a mediator decaying only to DM and a quark, with the natural width computed from Eq.~(\ref{Gammamin}). The black region in (a) is excluded from the pure QCD production of the mediator. Note that the mediator mass is denoted by $m_M$ in these plots instead of $\Mmed$ as in the rest of the text.}
\label{fig:t-channel1}
\end{figure}

\begin{figure}[t]
\begin{center}
\includegraphics[scale=0.5]{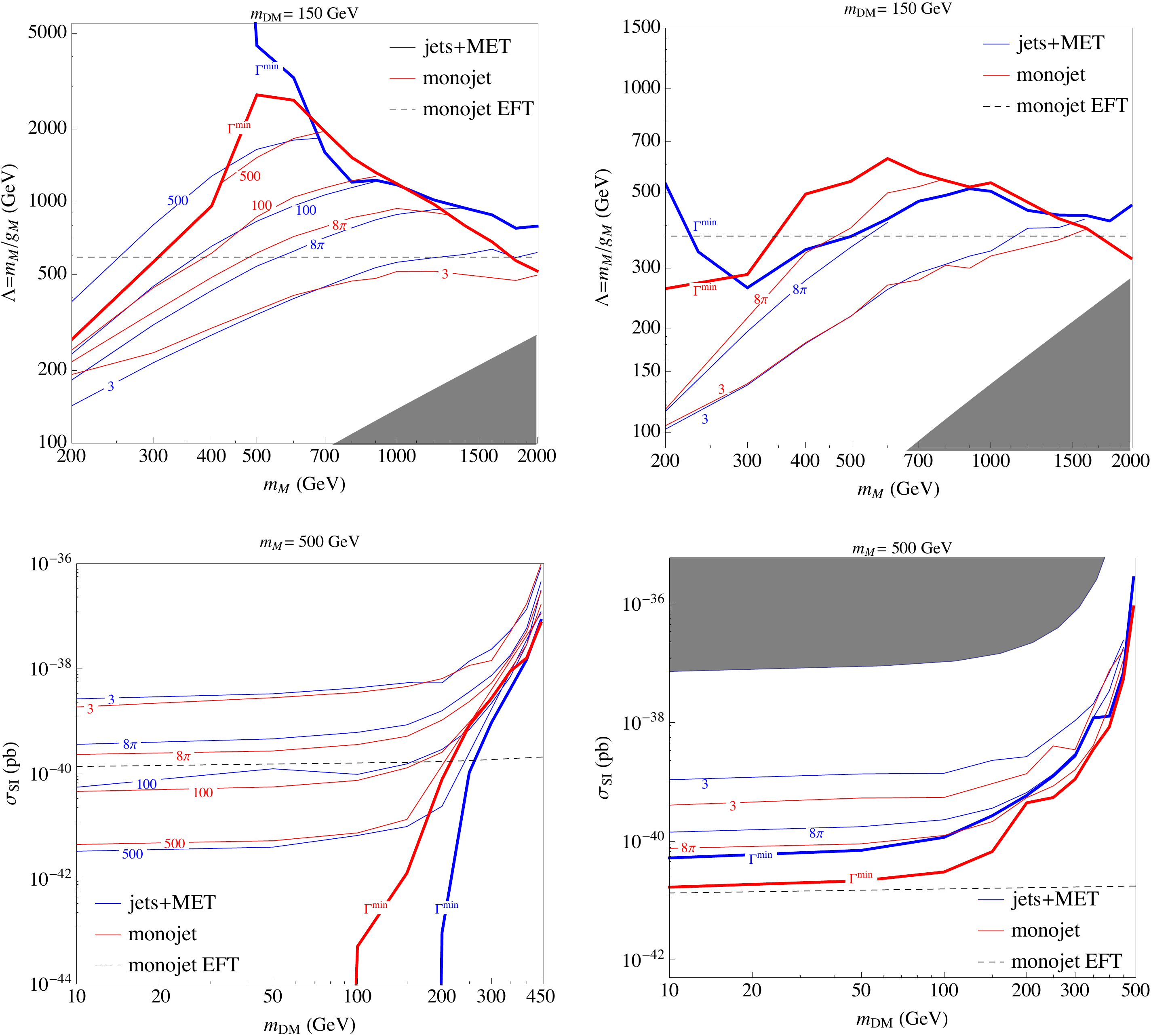}
\end{center}
\caption{Examples of exclusion limits on $\Lambda=\Mmed/g_M$ and direct detection cross section for two simplified models: $\widetilde{u},\widetilde{d},\widetilde{c},\widetilde{s}$, L+R (left column);  $\widetilde{d}_R,\widetilde{s}_R$ (right column).
Figures taken from ref.~\cite{Papucci:2014iwa}.  Note that the mediator mass is denoted by $m_M$ in these plots instead of $\Mmed$ as in the rest of the text.}
\label{fig:t-channel2}
\end{figure}

The parameter space for this colored-mediator + dark matter simplified model consists of three parameters only: the two masses $\Mmed$, $m_{DM}$ and the mediator-quark-dark matter coupling $g_M$. An intuitive and convenient way to visualize the results would be a color density plot in the $\Mmed$, $m_{DM}$ plane. In Figs.~\ref{fig:t-channel1},~\ref{fig:t-channel2} we report the exclusion bounds for the simplified model with all mediator flavors, $\widetilde{Q}_L^i, \widetilde{u}_R^i$, and $\widetilde{d}_R^i $. The model with only $\tilde d^i_R$-type mediators  can be treated in a similar manner and provides weaker constraints, owing to the smaller production cross section. From the plots we see that the interesting mass ranges are restricted to $100\text{GeV} \leq \Mmed \leq 2\text{TeV} $, $100\text{GeV} \leq m_{DM} \leq 1\text{TeV} $. Outside this region the sensitivity of the jets+missing energy searches decreases until eventually the interpretation of the model as a tree level exchange of a heavy resonance is lost, either because $g_{M}\sim 4\pi$ or because $\Gamma^{min}_{\rm med}(g_M)\sim \Mmed$.\footnote{Except in the compressed case regime, the latter happens before the former.}

\section{Simplified Models for $G_{\alpha\beta}G^{\alpha\beta}~\bar{\chi} \chi$ type operators}
\label{sec:ggchichi}

We now move on to consider the EFT operators associated with gluons in the initial state, such as the CP conserving operators,
\be
\frac{\alpha_s}{4\Lambda^3} {\rm tr}\left(G_{\mu\nu}G^{\mu\nu}\right) \bar{\chi}\chi, \quad {\rm and} \quad \frac{i\alpha_s}{4\Lambda^3} \epsilon^{\mu\nu\alpha\beta}{\rm tr}\left(G_{\mu\nu}G_{\alpha\beta}\right) \bar{\chi}\gamma_5\chi
\ee
where $\alpha_s$ is the strong coupling and $\Lambda$ denotes a high-energy scale. These are operators of the type D11-D14 of ref.~\cite{Goodman:2010ku}. Simplified models for such dimension-7 operators are more complicated.  In contrast to the simple resolutions we saw in sections~\ref{sec:qqchichi_s} and \ref{sec:qqchichi_t},  these operators cannot be resolved into two renormalizable operators glued by a single bosonic or fermionic mediators. Any resolution of these operators through a tree level exchange of a mediator will itself involve at least one non-renormalizable operator. Alternatively these operators can be resolved through a loop of mediators. We begin by considering the tree level resolution as it is sufficiently simple to be used as a simplified model. 

\begin{figure}[t]
\begin{center}
\includegraphics[width=0.9 \textwidth]{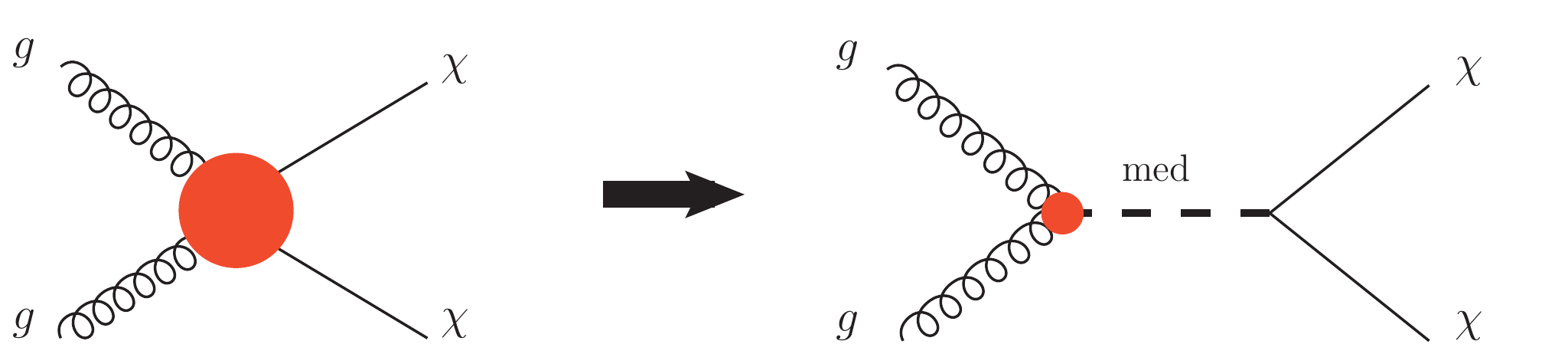}
\end{center}
\caption{The dimension-7 operators contributing to processes such as $gg\rightarrow \chi\chi$ on the left can be resolved to lowest order via the diagram on the right with a scalar in the $s$-channel. The red blob in the scalar-gluon vertex on the right serves as a reminder that this is also a non-renormalizable interaction that has to be resolved at some energy scale. }
\label{fig:simp_mod_gg_scalar_s_channel}
\end{figure}

A tree-level resolution of the dimension-7 operators is shown in Fig.~\ref{fig:simp_mod_gg_scalar_s_channel}: a scalar (or pseudoscalar) that couples through a renormalizable Yukawa-type interaction to the WIMP (a fermion) and with higgs-like (or axion-like) coupling to the gluons (dimension-5) is exchanged through the $s$-channel. The interactions take the form,
\be
\label{eqn:ggSchichi_int}
\mathcal{L} \supset  y_{_\chi} S\bar{\chi}\chi + \frac{\alpha_s}{\Lambda_{_S}} S G_{\alpha\beta}G^{\alpha\beta}
\ee
for the scalar and
\be
\label{eqn:ggSchichi_int2}
\mathcal{L} \supset  i y'_{_\chi} S'\bar{\chi}\gamma_5\chi + \frac{\alpha_s}{\Lambda_{_{S'}}} S' G_{\mu\nu}G_{\alpha\beta}\epsilon^{\mu\nu\alpha\beta}
\ee
for the pseudoscalar. Here $\Lambda_{_S}$ is some mass scale associated with the dimension-5 operator and the trace over the color indices has been left implicit.  The scale that appears in the dimension-7 operators (D11-14) is given parameterically by,
\be
\label{eqn:dim7_to_dim5}
\frac{\alpha_s}{4\Lambda^3} \sim \frac{1}{m_S^2}\frac{y_{_\chi}\alpha_s}{\Lambda_{_S}}
\ee
where $m_S$ is the mass of the scalar $S$. A similar expression holds for the case of a pseudoscalar or for multiple scalars. Current collider constraints on the dimension-7 operators from missing energy searches (see e.g.~\cite{ATLAS:2012ky}) give a bound of $\Lambda \gtrsim 350\GeV$. This is such a low scale that one must seriously wonder whether this description is valid at LHC energies. Indeed, as was shown in ref.~\cite{Busoni:2014sya} the EFT approach for this operator breaks down and the limit $\Lambda \gtrsim 350\GeV$ is invalid. 

Resolving the dimension-7 operator through a scalar or pseudoscalar exchange ameliorate this problem. As is clear from Eq.~(\ref{eqn:dim7_to_dim5}) by having a sufficiently light scalar (small $m_S$) we can have very low effective scale $\Lambda$. Importantly, this can be done consistently by keeping the dimension-5 scale, $\Lambda_{_S}$ sufficiently heavy to avoid any unitarity issues with this operator (at sufficiently high energies even this operator must be resolved as we discuss below). The scalars can now be produced on-shell and the $gg \rightarrow \chi\bar{\chi}$ process is dominated by this production. The interactions in Eqs.~(\ref{eqn:ggSchichi_int}) and (\ref{eqn:ggSchichi_int2}) can be easily implemented in existing event generators. In fact, the case of a scalar is entirely analogous to a heavy higgs boson that is produced on-shell through the usual gluon-fusion process and decays dominantly into missing energy.

Resolving the dimension-5 operator $S G_{\alpha\beta}G^{\alpha\beta}$ can be done if the scalar $S$ is coupled through Yukawa coupling to  some new heavy colored states (this is completely analogous to the Higgs coupling to gluons via the top quark loop). In the limit of heavy mediators' mass the dimension-5 coupling is related to the heavy colored states' mass and coupling through~\cite{Djouadi:2005gi}
\be
\label{eqn:higgs_type_coup}
\frac{\alpha_s}{\Lambda_{_S}} \propto \frac{\alpha_s}{8\pi}\sum_f \left( \frac{y_f}{M_f}\right)  
\ee
where the sum runs over all heavy colored fermions, $y_f$ is the Yukawa coupling of these fermions to the scalar $S$,  $M_f$ is the heavy fermions $f$. A similar expression holds for the case of a pseudo-scalar. So, this model can be resolved into a fully renormalizable model by introducing new heavy (vector-like) quarks that couple to the scalar mediator. The relations of Eq.~(\ref{eqn:higgs_type_coup}) and Eq.~(\ref{eqn:dim7_to_dim5}) require a mediator mass $m_{_S}$ which is not too heavy or the colored states are far too light and would have already been observed in searches for new colored states. 

\begin{figure}[t]
\begin{center}
\includegraphics[width=0.9 \textwidth]{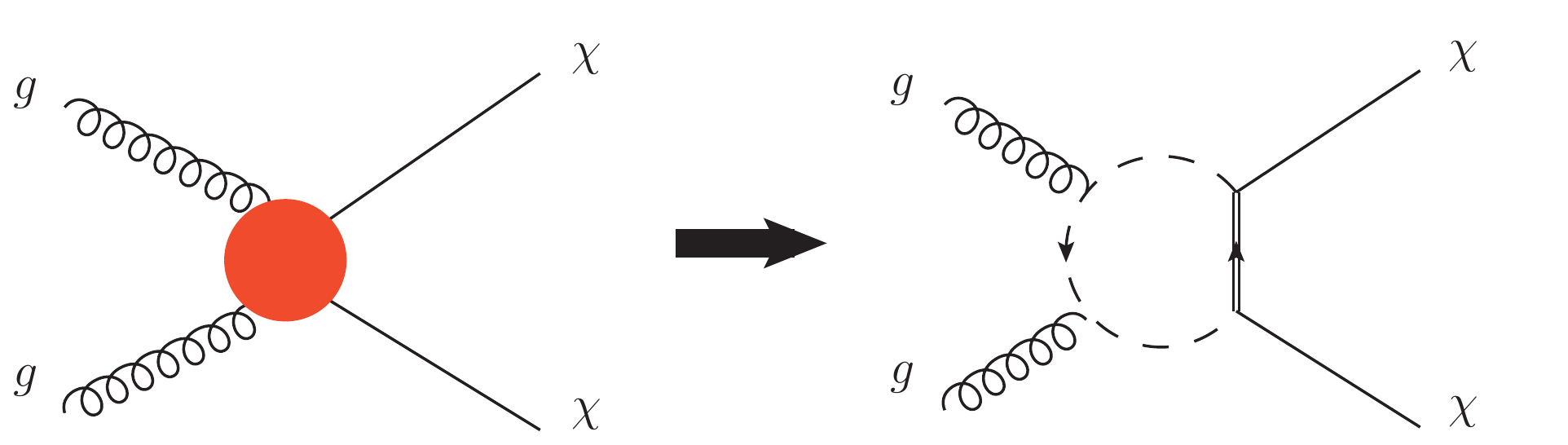}
\end{center}
\caption{The dimension-7 operator contributing to processes such as $gg\rightarrow \chi\bar{\chi}$ on the left is resolved on the right in a model with new colored scalars (dashed line)  and fermions (double line) that couple to the WIMP $\chi$.  Several other diagrams contribute aside from the one shown, see ref.~\cite{Weiner:2012gm} for details.}
\label{fig:simplified_model_gg_Rayleigh}
\end{figure}

It is also possible to resolve the dimension-7 operators directly into renormalizable interactions with colored mediators as was done for example in refs.~\cite{Weiner:2012gm,Frandsen:2012db} and is shown in Fig.~\ref{fig:simplified_model_gg_Rayleigh}. A simple example of such a model is one with new colored scalars and fermions that couple to the WIMP through a Yukawa-type interaction. The coupling of the dimension-7 operator is then related to the mass and coupling of these new states through,
\be
\frac{\alpha_s}{4\Lambda^3} \propto \frac{\alpha_s \lambda_\chi^2}{M_{\rm med}^3}
\ee
where $M_{\rm med}$ is the mass of the mediators and $\lambda_\chi$ is their coupling to the WIMP. Evidently, one needs fairly light mediators to generate the scale bounded by searches at the LHC, $\Lambda \sim 350\GeV$ as in ref.~\cite{ATLAS:2012ky}. Such new colored states are much easier to search for in other channels by producing them directly. Thus, this model is not very useful in providing a simplified framework to look for the process $gg\rightarrow \chi\bar{\chi}$. 

To conclude this section we reiterate that resolving dimension-7 operators of the type discussed above (D11-D14 of ref.~\cite{Goodman:2010ku}) in terms of simplified models is not as straightforward as it is for operators associated with quarks (e.g. D1-D10). Because of their high dimensionality using these EFT operators at the LHC is particularly problematic as was recently shown in~\cite{Busoni:2014sya}. Perhaps the simplest way of making sense of such operators is through a new higgs-like scalar (or pseudoscalar) that couples directly to the WIMP through a Yukawa coupling and to gluons through a dimension five operator as in Eqs.~(\ref{eqn:ggSchichi_int}) and~(\ref{eqn:ggSchichi_int2}).



\section{Simplified Models with Third Generation Quarks}
\label{sec:third_generation}

Models with dark matter coupled preferentially to third generation quarks have qualitatively different collider signals, including $b$-jets or higher multiplicity final states from top quarks. In the EFT approach, this corresponds to operators with flavor-dependent couplings. A flavor-safe way to treat this scenario is to assume minimal flavor violation, where the interaction strength for each flavor is proportional to the quark mass. The coefficients of the D1-D4 operators were parameterized to take this into account:
\begin{equation}
  {\cal O}_{D1}= \sum_q \frac{m_q}{\Lambda^3}\bar{q}q\bar{\chi}\chi,
\end{equation}
again assuming the DM is a fermion $\chi$. It is also straightforward to allow for different overall coefficients in the coupling to up-type and down-type quarks \cite{Kamenik:2011nb}.

\begin{figure}[tb]
\begin{center}
\includegraphics[scale=0.5]{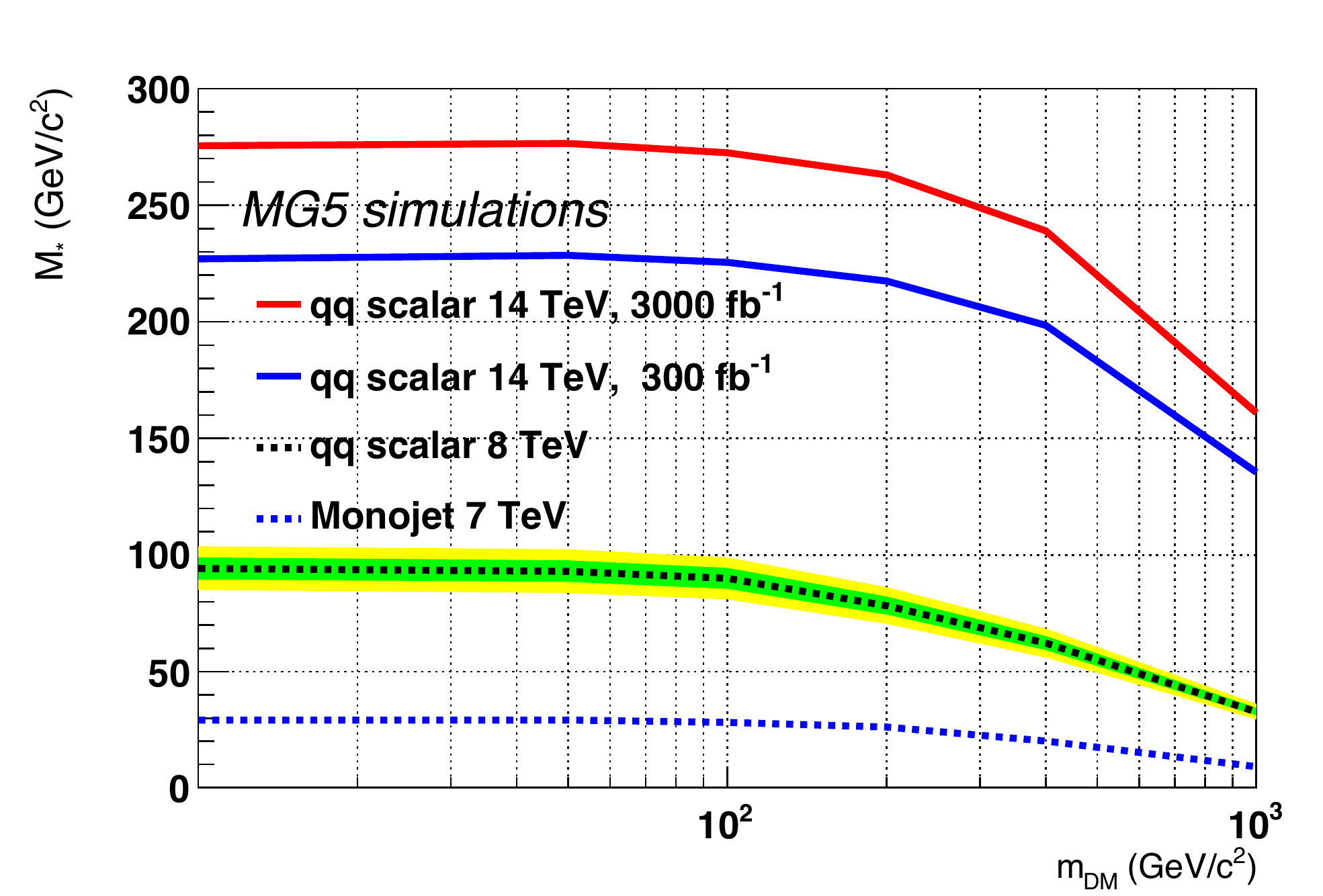}
\end{center}
\caption{Expected 90\% CL limits on the scalar operator D1 from a DM plus heavy quark search, including couplings to tops and bottoms.  ATLAS 7 TeV limits come from~\cite{ATLAS:2012ky}.}
\label{fig:combined_limits_mstar} 
\end{figure}

The enhanced couplings to heavy quarks for these operators led Ref.~\cite{Lin:2013sca} to consider the $b$-jet plus MET (mono-$b$) and $t \bar t$ plus MET collider signals. Despite the PDF suppression for producing these final states, it was found that limits could be improved significantly relative to the tree-level monojet limit. The irreducible background from $V$+jets was also reduced due to the requirement of a $b$-tag.
 Furthermore, Ref.~\cite{Haisch:2012kf} pointed out that  heavy quark loops lead to DM production through gluon fusion, which leads to much stronger monojet limits. However, in both cases the derived limit for light dark matter is $\Lambda \gtrsim 100$ GeV for the 8 TeV LHC run, in the case of the D1 operator as shown in Fig.~\ref{fig:combined_limits_mstar}. The limits are expected to increase to almost 300 GeV at 14 TeV~\cite{Artoni:2013zba}.

We briefly comment on the validity of the EFT assumption, following the discussion in Section.~\ref{sec:problems_with_EFT}.  Assuming an $s$-channel scalar mediator that couples primarily either to $b$-quarks or to top-quarks, the relation between the mediator mass $\Mmed$ and $\Lambda$ for the D1 operator above is given by:
\begin{equation}
	\Lambda =  \left( \frac{ m_q \Mmed ^2 }{ g_q g_\chi } \right)^{1/3}.
\end{equation}
For $\Lambda = 100$ GeV and coupling to $b$-quarks, the condition on the momentum transfer in Eq.~(\ref{conditionQM}) becomes $Q_{\rm tr}<  \sqrt{g_b g_\chi} \times 461$ GeV. Events passing mono-$b$ cuts can only satisfy this requirement for large couplings $\sqrt{g_b g_\chi}  \gtrsim 4$. The situation is worse for coupling to top quarks, which requires momentum transfer $Q_{\rm tr}<  \sqrt{g_t g_\chi} \times 76$ GeV. Even with extreme couplings of $4\pi$, the implied mediator mass is below 1 TeV, signaling the need for simplified models.

The simplest UV-complete possibility where DM couples to quarks proportional to their mass  arises in Higgs-mediated models, discussed further in Section~\ref{higgsportal}.  Simplified $s$-channel mediator models introduce a new neutral scalar \cite{Cotta:2013jna}, pseudoscalar or vector  \cite{Berlin:2014tja} analogous to those discussed in  Section~\ref{sec:qqchichi_s}, but the dominant interactions are with heavy quarks.  For example, Refs.~\cite{Izaguirre:2014vva,Ipek:2014gua} focused on a pseudoscalar, $a$, coupling primarily to $b$-quarks, which could arise through mixing with the pseudoscalar in a two-Higgs doublet model. 
The relevant interaction terms in this case are:
\begin{equation}
	{\cal L} \supset i (g_\chi \bar \chi \gamma^5 \chi + g_b \bar b \gamma^5 b) a
\end{equation}
It was shown in Ref.~\cite{Izaguirre:2014vva} that both the mono-$b$  and the sbottom search with two $b$-jets plus MET help to constrain the parameter space, depending on the $a$ mass.

Simplified $t$-channel models also have collider signals with heavy quarks, for example if there is a sbottom-like scalar mediator $\tilde B$:
\begin{equation}
\label{eqn:flavour_dm_coup}
	{\cal L} \supset - \lambda \tilde B \bar b_R \chi + \textrm{h.c.}
\end{equation}
which could arise in flavored dark matter models~\cite{Agrawal:2014una}. Note that these interactions do not necessarily generate the scalar (D1) operator above, and in addition the assumption of minimal flavor violation does not require interaction strengths to be proportional to mass. The sbottom search can be used to constrain the new mediator; however, when the coupling $\lambda$ in Eq.~(\ref{eqn:flavour_dm_coup}) is large additional channels open up relative to the SUSY case, which changes the final state kinematics.   Other $t$-channel models include instead DM coupling to the third generation left-handed doublet \cite{Chang:2013oia} or to right-handed top quarks \cite{Kumar:2013hfa,Batell:2013zwa}. 

In the presence of additional flavor-violating structure, single top plus MET (mono-top) production is possible 
 \cite{Kamenik:2011nb,Andrea:2011ws,Agram:2013wda}. An example is the simplified $t$-channel model of fermion dark matter coupled to top quarks, when the scalar mediator also has RPV-like couplings to light quarks. For a summary of the experimental signature, see Section~\ref{sec:exp_aspects}.

\section{Searches for the mediators}
\label{sec:mediators}

Aside from helping with the interpretation of missing energy searches in colliders, the simplified models we discussed above urge us to search for the mediators themselves. Indeed, it is precisely when the mediators are light that the simplified model approach is most needed. It is therefore only natural to consider searches for these mediators as part of a general program utilizing simplified models. This section outlines the main experimental signatures that can be used to constrain the 
parameter space of mediators coupling to quarks and gluons for a selection of the simplified models included in this report. Each of the searches mentioned includes a short description of the 
main experimental challenges, together with references to existing publication and reinterpretations
in terms of simplified models. 

\subsection{Vector mediator exchanged in the $s-$channel: $Z'$} 

The most common benchmark for the $s-$channel mediator is a color neutral vector boson ($\ZZ/Z'$). 
It can couple to fermionic DM particles either through an axial vector or a vector current,
and decay to various SM particles~\cite{Frandsen:2012rk, An:2012ue, An:2012va, deSimone:2014pda}. 

The mediator of DM/SM interactions could be the SM $\ZZ$ vector-boson. However, current searches at hadron colliders do not significantly add to the picture drawn by direct detection experiments and LEP precision constraints on the width of the $\ZZ$ boson~\cite{deSimone:2014pda}. Hadron colliders can on the other hand discover or constrain a leptophobic, high-mass $Z'$ that couples more strongly to the SM than to the DM sector~\footnote{For the case of an invisibly-decaying $Z'$, see Reference~\cite{Petriello:2008pu}.}. Such a $Z'$ would decay to quark-antiquark pairs, and appear as a resonant peak in the mass spectrum of central dijets. A $Z'$ of this kind is indeed used as a benchmark resonance in a variety of searches~\cite{Aaltonen:2008dn, ATLAS:2012pu, Chatrchyan:2013qha}, 
although it is generally assumed that it maintains the same couplings as the SM $\ZZ$~\cite{Altarelli:1989ff}. The hadronic decay of a new vector-boson could manifest itself in both the dijet mass distribution as well as the angular distribution, and is generally a well-motivated signature of new physics at hadron colliders~\cite{Harris:2011bh}. 

In the dijet mass resonance search, the overwhelmingly dominant QCD background 
is determined by fitting the data to a functional ansatz. This function will not accommodate deviations 
such as those introduced by a new resonant process. If no deviation is found between the data and the fit, 95\% CL limits are set on the cross sections of new physics models. The interpretation of the results of the dijet mass resonance search in terms of DM mediators is straightforward in the case of a narrow resonance for the ATLAS searches, using the limits provided in the Gaussian approximation for different values of the intrinsic resonance widths up to 15\%~\cite{Aad:2011fq}. Wider resonances will escape searches using a smooth function, instead appearing as an excess in centrally produced dijet events over the more forward, $t-$channel QCD background. The analysis of the ratio of central to total dijet events~\cite{Aad:2011aj} can be used for both wide resonance searches~\cite{An:2012va} and for contact interaction searches in the EFT framework~\cite{Dreiner:2013vla}.

The coupling between the mediator and both the WIMP ($\gDM$) as well as the SM ($\gSM$) come into play 
when considering the strengths of dijet searches as compared to the direct missing energy
searches for the WIMP.
If the coupling between the standard model and the mediator is large while the 
coupling of the dark sector to the mediator is weak, then the dijet search will generally
provide stronger bounds. On the other hand, a weak standard model coupling to 
the mediator and a strong dark sector coupling would be very hard to discover as 
a dijet resonance.  In particular, one can consider an example scenario 
of $\gSM=\frac{1}{4\pi}$ and $\gDM=4\pi$. 
Such a scenario would be challenging for dijet searches, while 
providing a reasonably strong signature in monojet topologies. 
A scan in the coupling-versus-mass plane as proposed in Reference~\cite{Dobrescu:2013cmh}
would further ease reinterpretation of the dijet resonance searches in terms of couplings.

Reference~\cite{Dobrescu:2013cmh} also points out an experimental shortcoming of recent dijet searches at colliders: 
constraints for low mass mediators are weakened by data taking limitations. The online resources devoted to recording and reconstructing the content of the collision events are finite and the high cross section of QCD backgrounds lead to only a small fraction of dijet events with masses below 1 TeV to be recorded. This in turn penalizes the statistical power
of searches and leaves unconstrained an interesting region of the mediators parameter space~\cite{Berlin:2014tja}. We therefore emphasize the importance of any possible progress in improving the reach of searches for new dijet resonances in the low mass region and their relevancy for DM searches. 

A new region of parameter space will be accessible with the 13 TeV center-of-mass energy dataset 
planned for the LHC Run-II. Sensitivity to TeV-scale resonances will be reached almost twice as fast
thanks to the increase in parton luminosities. However, the need for simplified models 
with high-mass mediators is reduced from the point of view of reinterpretations,
since in that region of the parameter space the EFT validity is not compromised~\cite{Buchmueller:2013dya}.
However, hints of a newly discovered high-mass resonance in the next LHC run 
will yield a wealth of questions on its nature, including its interpretation as a Dark Matter mediator. 

\subsection{Scalar mediator exchanged in the $s-$channel: Higgs portal} 
\label{higgsportal}

The use of simplified models to investigate and 
constrain specific mediators is of particular interest 
after the recent Higgs discovery. 
One topical question is whether the Higgs boson itself is a mediator 
between the standard model and a dark sector (Higgs Portal models~\cite{Fox:2011pm,Djouadi:2012zc,deSimone:2014pda}). 
Simplified models can be used to address this 
question by taking an $s-$channel scalar mediator
with the mediator mass set to the observed Higgs boson mass. 
By fixing the mediator mass, much more detailed 
studies in the $\gSM-\gDM$ coupling plane become feasible. 

Measurements of the properties of the SM Higgs boson 
are the main tool to constrain those models, 
albeit statistically and systematically limited at with present LHC data~\cite{Aad:2013wqa, CMS-PAS-HIG-13-005}.
Many manifestations of Higgs portal models would lead to a reduction or 
suppression of the Higgs boson couplings to SM particles,
in favor of its interactions with new particles~\cite{Englert:2011yb}.
Precision measurements of the Higgs couplings that can be undertaken
in future LHC phases and future accelerators can further constrain Higgs portal
models~\cite{Englert:2014uua}.

An alternative search strategy involves constraining the direct decays of the Higgs 
into WIMPs with $m_{DM} < m_H/2$. The Higgs partial width into invisible particles 
provides the key to a reinterpretation of those models in terms of cross
sections relevant for DM direct detection experiments~\cite{Fox:2011pm, Djouadi:2012zc}. The Higgs partial width can be experimentally constrained using
the Higgs branching ratios to invisible particles, in the vector-boson fusion 
and associated production channels~\cite{CMS-PAS-HIG-13-013},
\cite{Aad:2014iia}. Current ATLAS and CMS limits on invisible Higgs decay at the 95\% C.L. 
are around 70\%; they are expected to decrease to 20-30\% by the end of the upcoming $300\,\ifb$ LHC run
~\cite{ATL-PHYS-PUB-2013-015}. Finally, a recent recast of CMS search for stop quarks decays obtains a limit of about 40\% on the invisible branching ratio of the Higgs boson~\cite{Zhou:2014dba} . 

\subsection{Colored bosons and fermions exchanged in the $t-$channel} 

The presence of a light mediator exchanged in the $t-$channel leads to 
signatures with one or more jets. These signatures complement 
the monojet analyses where a jet from initial state radiation is exploited
to detect the presence of recoiling dark matter. This was discussed at length in section~\ref{sec:qqchichi_t} and here we only recap the qualitative features. 

In the $t-$channel case, the mediator can be either singly or pair produced. 
It will decay into a quark and a DM particle, and contribute 
to either single jet + MET or dijet + MET signatures.
SUSY squark searches in final states with high MET and 
two or more jets~\cite{ATLAS-CONF-2013-047, Chatrchyan:2014lfa} are sensitive to light mediators, 
and competitive with monojet searches depending on the masses of the mediator and DM particle. 
In the case of the monojet analysis, releasing the veto on the second jet allows sensitivity to 
pair production of mediators as well. The relative strength of each of the two search 
approaches for fermion and scalar mediators is compared in Refs.~\cite{Papucci:2014iwa, An:2013xka}, 
while fermion mediators are discussed further in ref.~\cite{Bai:2013iqa}.



\section{Experimental aspects of current and planned searches}
\label{sec:exp_aspects}

\def\gsim{\lower0.5ex\hbox{$\:\buildrel >\over\sim\:$}}
\def\lsim{\lower0.5ex\hbox{$\:\buildrel <\over\sim\:$}}
\def \rp{{R\hspace{-0.22cm}/}_P}
\def \lp{{L\!\!\!/}}
\def \n{\noindent}
\def\missET {{\not\!\! E_T}}

The LHC is a proton-proton collider with centre of mass energy of 7, 8 TeV which started data-taking at CERN in 2009 and will upgrade to higher centre of mass energies of 13-14 TeV in 2015. Two general-purpose experiments are installed at the LHC, ATLAS~\cite{atlas} and CMS~\cite{cms}. The energetic collisions taking place at the LHC have the potential for creating new particles not present in the SM and are naturally at the core of searches for new phenomenon. New particles with a lifetime that exceeds about a microsecond and which interact only weakly or sub-weakly with normal matter would escape detection and result in missing energy signatures. The elusive dark matter of the Universe is of course a prime example of such a particle, but it should be kept in mind that missing energy searches are much more inclusive and are sensitive to a wide variety of weakly interacting particles. Both ATLAS and CMS adopt the ``MET+$X$'' strategy to search for such invisible particles, which relies on the missing transverse energy ($\missET$) recoiling against an additional SM particle $X$ present in the process.

\subsection{MET+X searches}
The immense energies of the LHC motivate using many more SM particles as $X$ compared with what was previously used in past colliders. While traditionally missing energy searches were mostly done with $X$ as a photon or a jet (mono-photon and mono-jet searches), modern LHC searches utilize multijets, tops, $\ZZ$, $\WW$, the Higgs-boson, etc. While lower in rate, such experimental searches have the advantage of lower backgrounds. In addition to mono-jet~\cite{Aaltonen:2012jb, ATLAS:2012ky, Chatrchyan:2012me,Khachatryan:1750264} and mono-photon~\cite{Aad:2012fw,Chatrchyan:2012tea} searches, there are suggestions for mono-$Z$~\cite{Weiner:2012cb, Carpenter:2012rg, Bell:2012rg} and mono-$W$~\cite{Weiner:2012cb, Bai:2012xg}. By now both ATLAS and CMS have published dark matter search results with these channels in ref.~\cite{Aad:2014vka} (mono-$Z$), and refs.~\cite{Aad:2013oja, ATLAS:2014wra} (mono-$W$). There are many suggestions in the literature as well as some preliminary results for mono-top~\cite{Andrea:2011ws, Kumar:2013jgb, Agram:2013wda, CMS:2014hba} as well as proposals for mono-Higgs~\cite{Petrov:2013nia, Carpenter:2013xra, Berlin:2014cfa}. These different searches are all useful and important and should be pursued. Naturally, priority should be given to the broadest searches utilizing MET+jets. But, that should not be used as a reason not to also explore other MET+X searches as these represent orthogonal search directions covering models that can be missed with jets alone. As an example, we note that the associated production of a WIMP with a charged excited state which promptly decays to the WIMP through the emission of a $\WW$ can be efficiently searched for in the mono-$W$ channel~(see e.g.~\cite{Weiner:2012cb} for more details). As another simple example, the associated production of the Higgs with $\WW$ followed by the Higgs invisible decay leads to the same final state and can be searched for under the same channel. So, while searches for missing energy in conjunction with jets should continue with undiminished vigor,  searches utilizing other SM particles have their place in a healthy program looking for weakly interacting particles. 

\textbf{Mono-jet:} Interactions of quarks or gluons with WIMPs, such as the ones considered in previous sections, are most efficiently searched for using jets from initial state radiation. The mono-jet channel has the highest sensitivity in general due to the relatively large cross section~\cite{1302.3619}. In both ATLAS and CMS mono-jet searches, events were selected from $\missET$-related triggers, multiple signal regions were defined based on the $\missET$ and leading jet ($\missET, p_{\rm T}^{\rm leading}>$ 120, 220, 350, 500 GeV, $|\eta^{\rm leading}|<2.0$ at ATLAS and $\missET>$200, 250, 300, 350, 400 GeV, $p_{\rm T}^{\rm leading}>$110 GeV, $|\eta^{\rm leading}|<2.4$ at CMS). Only events with at most one additional jet of $p_{\rm T}$ above 30 GeV and no electron or muon candidates were selected. The dominant SM backgrounds here are $Z(\nu\nu)+$jets and $W+$jets where the lepton is not identified, which are estimated with data-driven methods based on $Z(\ell\ell)+$jets and $W(\ell\nu)+$jets events. Model independent limits were set for each signal region. The most stringent 90\% CL lower limits on the suppression scale for the D5 operator are around a TeV. Apart from multiple EFT operators, other signals such as ADD models of extra dimensions~\cite{ArkaniHamed:1998rs}, were tested. Expected sensitivity to DM in the mono-jet final states at 14 TeV is studied in ref.~\cite{ATL-PHYS-PUB-2014-007}. With a $\missET> 800$ GeV cut, expected lower exclusion limits at 2 TeV are foreseen for the D5 operator already after one year of data-taking.

\textbf{Mono-photon:} In the mono-photon analyses from 7~TeV data, ATLAS selected events with photon $E_{T}>150$~GeV, $|\eta|<2.37$, $\missET>150$~GeV and CMS selected events with photon $E_{T}>145$~GeV, $|\eta|<1.44$, $\missET>130$~GeV. Similar to the mono-jet channel, events with extra jets or leptons are vetoed. The dominant SM backgrounds are $Z(\nu\nu)+\gamma$ and $W+\gamma$ which were estimated from MC-simulation at CMS and data-driven method at ATLAS. There is non-negligible irreducible background coming from electron's misreconstruction as photon. Lower limits in the suppression scale at 90\% CL are at the order of 10 GeV for the D1 operator and close to 600 GeV for the D5 operator. Upper limits are then placed on the dark matter production cross sections assuming the same EFT model as the mono-jet analyses.  

\textbf{Mono-$W$, Mono-$Z$:} For the mono-$Z$ and mono-$W$ analyses, searches for dark matter signals in both leptonic and hadronic decay channels were conducted in 8 TeV data collected by ATLAS~\cite{Aad:2014vka, Aad:2013oja, ATLAS:2014wra}. In the hadronic channel, events are selected with $\missET$-related trigger and offline $\missET$ cuts ($\missET>350, 500$~GeV). Therefore the balanced $Z$ or $W$ bosons are produced with a strong boost. These boosted $W$ and $Z$ can be reconstructed as a large-radius jet ($p_{T}>250$~GeV, $|\eta|<1.2$) and discriminated from QCD jets through jet substructure information. Given that they have the same final states, mono-$W$ and mono-$Z$ channels are combined together to improve the sensitivity. The dominant SM backgrounds in this hadronic channel are $Z(\nu\nu)+$jets and $W+$jets which are estimated from events containing electrons or muons using data-driven methods similar to mono-jet. 
The mono-$W$ leptonic analysis combines both electron and muon decay channels and optimizes the signal regions with multiple cuts on the transverse mass of the lepton-neutrino system. The SM background is dominated by $W+$jets. The mono-$Z$ leptonic channel combines both electron and muon decays with signal regions $\missET>$150, 250, 350, 450~GeV. The dominant SM process $ZZ$ is derived from MC simulation. 
These mono-boson (photon, $Z$, $W$) channels are interesting due to the fact that they can probe the dark matter couplings to bosons. For instance, ATLAS mono-$Z$ analysis tested a specific EFT model with $Z$ produced directly from the dark matter production vertex. Furthermore, the mono-W analyses are sensitive to the differences in the couplings of DM to u- and d-quarks due to the interference of the diagrams involved. The extreme cases result in an order of magnitude difference in the WIMP-nucleon cross section. The inferred 90\% CL limit on the suppression scale of the D5 operator in the constructive interference mode is around 2 TeV and it is the most stringent constraint among all mono-X searches. Constraints from these channels can be converted into limits on dark matter annihilation into bosons, which are particularly relevant for indirect searches for DM through its present-day annihilation in the galaxy. In addition, the mono-$Z$ is one of the channels with the highest sensitivity to Higgs-portal DM models (e.g.\ refs.~\cite{Burgess:2000yq,Cline:2013gha}) where the Higgs is produced in association with a $Z$-boson and consequently decays invisibly into the DM candidates. 

\textbf{Mono-top, Mono-$t\bar{t}$:} CMS have some preliminary results for dark matter in association with single top~\cite{CMS:2014hba} and top pair~\cite{CMS:2014pvf}. These searches are motivated by specific models where dark matter has large couplings with heavy quarks as in ref.~\cite{Lin:2013sca,Andrea:2011ws}. The mono-top search focuses on top hadronic decays and selects events with large $\missET$ ($\missET>350$~GeV) and three jets consistent with a top decay. Similar to the case of the monojet analysis, a lepton veto requirement is implemented here as well. The dominant backgrounds are $Z+$jets and $t\bar{t}$. The mono-$t\bar{t}$ search selects $t\bar{t}$ dilepton decay events with $\missET>320$~GeV. Other kinematics cuts based on the two leptons are also applied to enhance the signal sensitivity like dilepton invariant mass and opening angle in transverse plane. The 90\% CL limit on the suppression scale of the D1 operator is set around 100 GeV.

Many mono-$X$ searches rely on the $\missET$ trigger which indicates that the offline $\missET$ selection cannot be lower than $\sim 100$~GeV. This threshold is supposed to increase for LHC Run II with higher energy and luminosity. Signals with weak boost like Higgs-portal model may suffer from this trigger threshold. 

Given the broad nature of MET+X searches, they can be, and indeed have been, interpreted in a variety of frameworks: from EFTs, through Simplified Models, to full fledged models such as SUSY, ADD, etc. We endorse this pluralism and encourage experimenters to remain cognizant of these different points of view as they work to bolster existing effort as well as explore new grounds.


\section{Comments and recommendations regarding the use of simplified models}
\label{sec:recommendations}

Many models of new physics provide dark matter candidates, either by design or as a side effect. For UV-complete, fully specified models, one can assess the complete set of possible signatures of a given model and design a targeted search program. Such targeted searches can maximize the chances of discovery, and often involve additional event topologies as compared with the ones considered in this review. But despite the imagination and hard work that goes into such specific models, they are only a finite set and the true theory of dark matter interactions may not be among them. When the model is unknown, general purpose and broad searches should also be employed.

The previous sections have argued that simplified models of WIMP pair production are one important tool for designing and interpreting broad and inclusive searches at colliders. These simplified models each contain a single species of WIMP and a single interaction between it and the Standard Model, represented by a new mediating particle that might be directly observed at LHC energies. Taken together, the set presented above can be considered likely building blocks out of which any proper theory involving interactions between the SM and the dark sector would be constructed. Nevertheless, it is important to keep the limitations of simplified models firmly in mind. In this section, we discuss how they should---and should not---be used to maximize the usefulness of experimental results.

\subsection{Designing a search program}
\label{designingasearchprogram}

We believe that the simplified models discussed in this report represent a very broad class of models, and certainly include the best motivated examples of new physics that we are aware of.  However, given the immense importance of this search program, we now briefly discuss some important caveats to keep in mind. 

It is important to remember that simplified models are incomplete models. The full theory may be, like the Standard Model, more complex than a single dark matter particle with a single interaction. The processes relevant to colliders may not necessarily be the same as the processes relevant to WIMP-nucleon scattering in direct detection, nor must they be the same as the WIMP pair annihilation processes relevant to indirect detection experiments. Moreover, the real theory may contain multiple mediators, one mediator important for one search, and another type of mediator for a different search. This has important implications:

\begin{itemize}
\item Before attempting combinations across different channels, one should carefully check that the mediator and its couplings are the same in each channel. Otherwise, one is risking throwing away information based on what may very well be an incorrect assumption. Instead, as much as possible results should be quoted for each channel separately. Combinations can be done separately and the assumptions that go into such combinations should be clearly spelled out.  

\item When planning a search, care must be taken that prior collider, direct detection, and indirect detection constraints are considered only when directly applicable to a given operator.

\item The simplified model approach should not be used only to reinterpret familiar EFT-focused searches in terms of light mediators. It should also inform us whether additional search channels are necessary, particularly in cases where the mediator coupling to visible particles produces qualitatively different final states, or when the mediator is very light. In the full theory, it is possible that more than one of the processes described by the simplified models is actually relevant to collider searches, so all of the building blocks must be covered.
\end{itemize}

From the experimental point of view, it is important to make sure that the
largest possible part of the phase space is searched for physics beyond the
Standard Model. The EFT-focused searches are generic in the sense that they do not
involve any complicated, model-specific topologies. 
In a sense, such searches only investigate very simple final states like
mono-X. When going beyond EFT, it is important to choose which extensions of the
simple mono-X topologies may also lead to significant improvements in
sensitivity to new signals. These questions are already model dependent. In the
end, the results for a chosen set of models will be given only in a limited set
of the available LHC analyses. It is worth mentioning that a generic, truly
model independent search for new physics exists that monitors data to Standard Model Monte
Carlo agreements in dozens of different final states involving various
combinations of final states~\cite{ATLAS-CONF-2014-006}.
Such a search was performed in ATLAS once most other searches
were completed, in order to ensure that no signal was missed. 

\subsection{Recommendations for simulation}
\label{recommendationsforsimulation}

For the reinterpretation of the LHC results, it is important to clearly specify
how the signal Monte Carlo samples are generated. For example, it is not
complicated to implement new models into MadGraph using for example FeynRules.
However, details such as matching methods for parton showering and PDF set used may lead to
different topologies and kinematical distributions in the end.
Details of the Monte Carlo production should be made clear so that anyone can
reproduce the signal samples for reinterpretation\footnote{We note that being able to reproduce the signal used in an analysis is an important step in any reinterpretation effort as a way to build confidence in the reproduction of the analysis.}.

Using next-to-leading order (NLO) results for the production cross-section of a signal associated with a particular model is reasonable, and might be especially important if the effect is not simply a rescaling of the overall cross-section but rather results in a change in the momentum distributions for example. However, at this point in time the case for inclusion of NLO corrections is not very strong. For example, Ref.~\cite{Haisch:2013ata} investigated NLO effects for $s$-channel processes and found that the corrections to the inferred limits are inconsequential (although the theory uncertainties are reduced). Given that NLO results involve greater complexity and generally require more work to reproduce, we recommend that they be implemented only when a strong case for their inclusion is made and their effects are shown to be significant. 
If, and when additional complexity is deemed necessary, we urge the experimental collaborations to make sure that enough details are provided to allow for a full and, as much as possible, straightforward reproduction of the signal used in the analysis. Similar comments apply to the inclusion of uncertainties associated with parton distribution functions.

\subsection{Recommendations for the presentation of results}
\label{recommendationsforthepresentationofresults}

Experimental results addressing simplified models are most useful if they include all information needed to reinterpret the results in the context of a larger theory. A good starting point is the Les Houches Recommendations for the Presentation of LHC Results~\cite{Kraml:2012sg}. At minimum, the results should provide the observed event yields along with the predicted SM background and uncertainty, in inclusive bins of the key kinematic variables (e.g. the transverse momentum, $p_T$, and rapidity, $\eta$, of the jet in monojet searches). Results should provide as much numerical detail as possible (e.g. histograms) in auxiliary repositories such as HEPdata. Most importantly, as mentioned above, the results should be quoted separately for individual channels. In the future it might be interesting to try and implement the experimental analysis into a framework like RECAST~\cite{Cranmer:2010hk} to allow for streamlined reinterpretation of the results by different members of the theoretical and experimental communities. 

The set of free parameters in these models is small. For example, the $s$-channel resonance production of DM discussed in section~\ref{sec:qqchichi_s} depends on: the dark matter mass ($m_{\rm DM}$); the mediator mass ($M_{\rm med}$); and the product of the couplings of the mediator to the dark mater ($\gDM$) and the SM ($\gSM$). It is also possible to include the width of the resonance as it does affect the phenomenology when it is large, as discussed in section 4.1 of ref.~\cite{Buchmueller:2013dya}. As is common practice for simplified models in other contexts, limits should be provided as a function of all of the free parameters, rather than making unwarranted assumptions about them. An example of such a plot is shown in our Fig.~\ref{fig:t-channel1} taken from ref.~\cite{Papucci:2014iwa}. Another example is Fig. 8 of the ATLAS mono-Z search paper~\cite{Aad:2014vka} which provides the upper limit on the mediator-WIMP coupling as a simultaneous function of both the mediator mass and the mass of the WIMP. Similar figures should be included in all such experimental results.

At the same time, results employing the EFT can still be useful. The validity conditions described above can have a significant effect on them. But, provided this effect is properly quantified, they remain the most generic description available of the possible low-energy phenomenology. The range of validity of the EFT can be addressed within experimental papers, when presenting the limits on the EFT suppression scale and when comparing nucleon-WIMP cross section limits to Direct Detection experiments. Two suggestions can be found in References~\cite{Busoni:2014sya, Berlin:2014cfa}, in Fig.~8 and Fig.~3 respectively. Both examples involve limiting the momentum transfer in the signal generation to ensure the production of meaningful events within the EFT framework. 

One convenient aspect of the EFT approach is the ease with which collider constraints can be applied to the phenomenology of direct detection experiments (see e.g.\ ref.~\cite{Goodman:2010ku}). The same reinterpretation can be made using simplified models. For example, consider the case of $s$-channel mediator discussed in section~\ref{sec:qqchichi_s}. Collider searches place limits on the mass and couplings of the mediator through its $s$-channel contribution to the production of DM. The same interactions would show up in direct detection experiment in the scattering of DM against the nucleus through the mediator exchange. The rate in direct detection experiments depends on precisely the same set of parameters as those constrained by collider searches (namely the mass and couplings). Thus, it is still possible to make inferences about direct detection experiments from constraints on simplified models in colliders. 

When the mass of the mediator is above the typical momentum exchange in direct detection experiments (a few MeV),  the interaction is essentially a contact interaction. The simplified model can then just be mapped to the corresponding EFT operator in order to derive the bound in direct detection experiments. To be clear, the simplified model is used to place a limit using collider constraints where the energies are sufficiently high that the interactions between DM and the partons cannot be treated as an EFT operator. But, then the constraints on the coupling and mass of the mediator can be combined to yield a constraint on the EFT scale, $\Lambda$, which in turn can be used to set limits on the rate in direct detection experiments where the energies are sufficiently low to allow for "integrating-out" the mediator. So for example, in the case of an $s$-channel vector mediator with vector-like couplings, the mass $M_{\rm med}$ and coupling $g_q$ and $g_\chi$ combine to give a bound the scale $\Lambda = M_{\rm med}/\sqrt{g_q g_\chi}$ of the operator D5. 

This procedure is of course invalid once the mediator mass drops below a few MeV. At this point the interaction is no longer contact like even in direct detection experiments. The actual cross-section does not continue to rise as the mediator mass is lowered, instead it saturates at a mediator mass of around an MeV corresponding to the typical momentum exchange. We recommend that this case is left for theorists at this stage and the experimental collaboration simply presents results only above an MeV or so. This does not take away any power from the results as anyone interested in the region below an MeV can use the direct constraints on the couplings to work out the precise bounds on direct detection in this low mass region.

Finally, when drawing conclusions from searches using both simplified models and EFT frameworks, it is important to remember that they are incomplete. The set of simplified models above is a set of building blocks out of which a proper theory can be constructed. Collider searches can convey most of the information available in X+MET searches by constraining the building blocks. Constraints on proper models, models that are theoretically sound, free of anomalies, etc., should be left to theorists.


\section{Conclusions}
\label{sec:conclusions}

This present work was devoted to the use of simplified models of dark matter in missing energy searches at the LHC. The need for simplified models stems from the desire to interpret missing energy searches in as model-independent fashion as possible, and yet avoid the problems associated with the EFT approach (Sec.~\ref{sec:problems_with_EFT}). We presented and discussed a minimal set of different simplified models needed for general DM searches and their reinterpretation. In addition to the DM candidate, these models contain another new particle referred to as a ``mediator" which mediates the interactions of the SM with the DM. Particular attention was given to models where the mediator leads to DM production from quark - antiquark annihilation through the $s$-channel (Sec.~\ref{sec:qqchichi_s}) and $t$-channel (Sec.~\ref{sec:qqchichi_t}). We also proposed models where the production of DM is through gluon fusion (Sec.~\ref{sec:ggchichi}) and considered the subtleties involved with the corresponding operators. Finally, we reviewed some of the recent developments associated with models where DM interacts mostly with bottom and top quarks (Sec.~\ref{sec:third_generation}). 

When the mediator is sufficiently light to be produced on-shell other processes aside from DM production must be considered. Such processes can be (and are) searched for directly and are an important part of a comprehensive program to search for DM in colliders (Sec.~\ref{sec:mediators}). In addition to reviewing the experimental aspects of traditional searches for missing energy in association with jets or photons, we also briefly discussed more recent proposals to use other objects such as mono-W/Z, mono-top, and other possibilities (Sec.~\ref{sec:exp_aspects}). Finally, we put forward a set of recommendations for the design of future searches in Run-II of the LHC including the simulation of simplified models as well as the presentation of results and their reinterpretation as constraints on signals at direct-detection experiments (Sec.~\ref{sec:recommendations}). We hope the present work will aid in future efforts searching for new physics associated with DM in colliders.

\acknowledgments{TGR work was funded under the Department of Energy, Contract DE-AC02-76SF00515. SS is supported by funds from the Natural Sciences and Engineering Research Council (NSERC) of Canada and by the W. Garfield Weston Foundation. TV and EE are supported in part by the Israel Science Foundation, the US-Israel Binational Science Foundation,  and by the I-CORE Program of the Planning Budgeting Committee and the Israel Science Foundation (grant NO 1937/12). TV is also supported by the EU-FP7 Marie Curie, CIG fellowship. IY is supported in part by funds from the NSERC of Canada. This research was supported in part by Perimeter Institute for Theoretical Physics.}
\newpage
\clearpage

\bibliography{bibliography}

\end{document}